\documentclass[aps,prd,amssymb,eqsecnum,showpacs]{revtex4}
\usepackage{graphicx}
\usepackage{psfrag}

\usepackage{graphicx}
\usepackage{psfrag}

\begin{document}

\title{Harmonic Initial-Boundary Evolution in General Relativity}

\author{Maria C. Babiuc${}^{1}$ ,
	B\'{e}la Szil\'{a}gyi${}^{2}$ ,
	Jeffrey Winicour${}^{1,2}$
       }
\affiliation{
${}^{1}$ Department of Physics and Astronomy \\
         University of Pittsburgh, Pittsburgh, PA 15260, USA \\
${}^{2}$ Max-Planck-Institut f\" ur
         Gravitationsphysik, Albert-Einstein-Institut, \\
	 14476 Golm, Germany
	 }

\begin{abstract}

Computational techniques which establish the stability of an evolution-boundary
algorithm for a  model wave equation with shift are incorporated  into a
well-posed version of the initial-boundary value problem for gravitational
theory in harmonic coordinates. The resulting algorithm is implemented as a
3-dimensional numerical code which we demonstrate to provide stable, convergent
Cauchy evolution in gauge wave and shifted gauge wave testbeds. Code
performance is compared for Dirichlet, Neumann and Sommerfeld boundary
conditions and for boundary conditions which explicitly incorporate constraint
preservation. The results are used to assess strategies for obtaining
physically realistic boundary data by means of Cauchy-characteristic matching.

\end{abstract}

\pacs{PACS number(s): 04.20Ex, 04.25Dm, 04.25Nx, 04.70Bw}

\maketitle

\section{Introduction}

The ability to compute the details of the gravitational radiation produced by
compact astrophysical sources, such as coalescing black holes, is of major
importance to the success of gravitational wave astronomy. The simulation of
such systems by the numerical evolution of solutions to Einstein's equations
requires a valid treatment of the outer boundary. This involves the
mathematical issue of proper boundary conditions which ensure  a well-posed
initial-boundary value problem (IBVP), the computational issues of the 
consistency, accuracy and stability of the finite difference approximation, and
the physical issue of prescribing boundary data which is free of spurious
incoming radiation. The physical issue can be solved by locating the outer
boundary at future null infinity ${\cal I}^+$ in a conformally compactified
spacetime~\cite{penrose}. Since no light rays can enter the spacetime
through ${\cal I}^+$, no boundary data are needed to evolve the interior
spacetime. In addition, the waveform and polarization of the outgoing
radiation can be unambiguously calculated at ${\cal I}^+$ in terms of the
Bondi news function. This global approach has been applied to single black
hole spacetimes using Cauchy evolution on hyperboloidal time slices
(see~\cite{Hhprbloid,joerg} for reviews) and using characteristic evolution
on null hypersurfaces(see~\cite{winrev} for a review). However, most current
computational work on the binary black hole problem involves Cauchy
evolution in a domain whose outer boundary is a finite timelike worldtube. 
This approach could be justified physically if the boundary data were
supplied by matching to an exterior solution extending to ${\cal I}^+$. The
present work is part of a Cauchy-characteristic matching (CCM)
project~\cite{vishu,harl} to achieve this by matching the interior Cauchy
evolution to an exterior characteristic evolution. The success of CCM
depends upon the proper mathematical and computational treatment of the
Cauchy IBVP. A key issue, on which we focus in this paper, is the  accurate
preservation of the constraints in the treatment of the Cauchy boundary. The
results have important bearing on the matching problem. 

Early computational work in general relativity focused on the initial value
problem in which data on a spacelike hypersurface ${\cal S}$ determines the
evolution in its domain of dependence. However, in the simulation of an
isolated system, such as a neutron star or black hole, ${\cal S}$ typically
has an outer timelike boundary ${\cal B}$, coincident with the boundary of
the computational grid. The initial-boundary value problem addresses the
extension of the evolution into the domain of dependence of ${\cal
S}\cup{\cal B}$.  The IBVP for Einstein's equations is still not well
understood due to complications arising from the constraint equations. 
See~\cite{friedrend} for a  review of the mathematical aspects.
Well-posedness of the mathematical problem, which guarantees the existence
and uniqueness of a solution with continuous dependence on the Cauchy and
boundary data, is a necessary condition for computational success. A
well-posed IBVP for a linearization of the Einstein equations was first
presented by Stewart~\cite{stewartbc}, and the first well-posed version for
the full nonlinear theory was established by Friedrich and
Nagy~\cite{Friedrich98}. These works have influenced numerous
investigations of the implementation of boundary conditions in numerical
relativity~\cite{bishbc,szilschbc,harl,callehtigbc,calpulreulbc,fritgombc,
fritgombc3,bab,fritgombc4,sartig,gunmgarcbc,caltechbc,bonabc,multiblock,
calabgund,linsch}.
However, the combined mathematical and computational aspects of the IBVP still
present a major impasse in carrying out the long term simulations necessary to
compute useful waveforms from the inspiral and merger of black holes.

The IBVP for general relativity takes on one of its simplest forms in a
harmonic gauge, in which the well-posedness of the Cauchy problem was first
established~\cite{Choquet}. In previous work, we formulated a well-posed constraint
preserving version of the harmonic IBVP and implemented it as a numerical
code~\cite{harl}, the Abigel code. The IBVP was demonstrated to be well-posed for
homogeneous boundary data (or for small boundary data in the sense of
linearization off homogeneous data) and the code was shown to be stable and
convergent. Nevertheless, numerical evolution was limited by the excitation of
exponential instabilities of an analytic origin. Insight into the nature of
these unstable modes and mathematical and computational techniques for dealing
with them have been developed in a study of the evolution problem in the
absence of boundaries. These studies were carried out on toroidal spatial
manifolds, equivalent to the imposition of periodic boundary conditions. In this
paper, we apply the techniques which were successful in the periodic case to
the general harmonic IBVP. Although our applications here are limited to test
problems, we expect the techniques will be of benefit in furthering recent
progress in the simulation of black holes by harmonic
evolution~\cite{pret1,pret2,linsch}. 

Our previous work on the IBVP~\cite{harl} was based upon a reduced form of the
harmonic system due to Fock~\cite{Fock}. We present our results here in the
more  standard form~\cite{wald,friedrend} used in most analytic work. We have
described and tested a finite difference evolution code based upon this
standard harmonic formulation in~\cite{babev}. In Sec.~\ref{sec:cauchy}, we
summarize the main analytic and computational features of this evolution code.
The formalism  includes harmonic gauge forcing terms and constraint
adjustments. In principle, gauge forcing terms~\cite{Friedrich} allow the
simulation of any nonsingular spacetime region. Gauge forcing not only allow
the flexibility to ``steer'' around pathologies that might otherwise arise in
standard harmonic coordinates but it also allows universal adaptability to
carry out standardized tests for code performance, such as the AppleswithApples
(AwA) tests~\cite{mex1}. The formalism also includes constraint adjustments
which modify the nonlinear terms in the standard harmonic equations by mixing
in the harmonic constraints. Such constraint adjustments have proved to be
important in harmonic evolution in suppressing instabilities in a shifted
version of the AwA gauge wave test~\cite{babev} and in simulating
black holes~\cite{pret1,pret2}.

Constraint adjustments, when combined with a flux conservative form of the
equations, can lead to conserved quantities which suppress exponentially
growing error modes. As a dramatic example, the AwA gauge wave tests for the
Abigel harmonic code show an increase in error of more than 12 orders of
magnitude, after 100 grid crossing times, if flux conservation is not
employed~\cite{bab}. The error arises from exponentially growing solutions to
the reduced equations which have long wavelength and satisfy the constraints,
so that it is unaffected by either numerical dissipation or constraint
adjustment. The underlying conservation laws only apply to the prinicple part
of the system. The are effective when the nonlinear terms corresponding to the
unstable modes are small, or can be adjusted to be small by mixing in the
constraints. This is the case with the AwA gauge wave. However, as we discuss
in Sec.~\ref{sec:cauchy}, a shifted version of the gauge wave test excites a
different type of exponential instability which does not satisfy the
constraints and whose suppression requires constraint adjustment in addition to
flux conservation. This raises the caveat that the identification of the
nonlinear instabilities in the system is critical to the effectiveness of
constraint adjustment and conservative techniques.

In Sec's \ref{sec:hom} and \ref{sec:inhom}, we describe how our earlier
proof~\cite{harl} of the well-posedness of the harmonic IBVP for Einstein's
equations extends to the generalized formulation considered here. Constraint
preservation imposes requirements coupling the intrinsic metric and extrinsic
curvature of the boundary, analogous to the momentum constraint on Cauchy data.
However, the boundary data has fewer degrees of freedom than the corresponding
Cauchy data, as for a scalar field where either the field or its normal
derivative, but not both, can be specified on the boundary. This couples the
evolution to the specification of constraint-preserving boundary data in a way
that the standard theorems used to establish well-posedness only apply to
boundary data linearized off homogeneous boundary data. In the case of large
boundary data, one is only assured that the solution, if it exists, does
satisfy the constraints.

In Sec.~\ref{sec:imp}, we present the techniques used to implement the formalism
as a finite difference evolution-boundary algorithm. Since the preliminary
testing of the Abigel code, considerable improvement has been made in the
numerical techniques. The long term performance of the evolution algorithm in
the AwA gauge wave test with periodic boundaries has been improved
by use of semi-discrete conservation laws~\cite{bab}. The study of a model
nonlinear scalar wave~\cite{excis} shows how these semi-discrete conservation
laws can be used to formulate stable algorithms for more general boundary
conditions.

In Sec.~\ref{sec:tests}, we demonstrate the stability and second order
convergence of the code for periodic, Neumann, Dirichlet and Sommerfeld
boundary conditions. These tests are performed using gauge wave and shifted
gauge wave metrics for which the exact solutions provide the correct boundary
data. In principle, the knowledge of the exact boundary data avoids the need
for constraint-preserving boundary conditions. However, numerical noise can
generate constraint violating error. This is investigated by comparing these
constraint free boundary algorithms against the constraint-preserving algorithm
discussed in Sec's~\ref{sec:hom} and \ref{sec:inhom}. In order to eliminate the
complication of sharp boundary points, the tests are carried out by
opening up the 3-torus (periodic boundary) into a 2-torus times a line, with
two smooth 2-toroidal boundaries. 

Just as the ``3+1'' decomposition $x^\mu=(t,x^i)$ is useful in describing the
geometry of a spacelike Cauchy hypersurface ${\cal S}$ at $t=0$, the 
decomposition $x^\mu=(x^a,x)$, with $x^a=(t,y,z)$ is useful in describing the
geometry of a timelike boundary ${\cal B}$ at $x=0$. We distinguish between
these decompositions by using indices $i,j,k,...$ near the middle of the
alphabet or indices $a,b,c,...$ at the beginning of the alphabet. In order to
avoid excessive notation we indicate the corresponding intrinsic 3-metrics by
$h_{ij}$ and $h_{ab}$, with inverses $h^{ij}$ and $h^{ab}$. Where this might
cause confusion, we introduce subscripts, e.g. $h_{\cal S}=det(h_{ij})$ and
$h_{\cal B}=det(h_{ab})$. Curvature
conventions follow~\cite{wald}. This introduces some sign changes from the
treatment in~\cite{harl} which was based upon the conventions in~\cite{Fock}.
We use the shorthand notation $\partial_\alpha f =f_{,\alpha}$ where confusion
does not arise.

\section{The generalized harmonic Cauchy problem}
\label{sec:cauchy}

We summarize here the treatment of the generalized harmonic Cauchy problem on
which the evolution code described in~\cite{babev} is based. Generalized
harmonic coordinates $x^\alpha=(t,x^i)=(t,x,y,z)$ are functionally independent
solutions of the curved space scalar wave equation,
\begin{equation}
    \Box x^\mu  = \frac{1}{\sqrt{-g}}\partial_\alpha
           (\sqrt{-g}g^{\alpha\beta}\partial_\beta x^\mu) =-\hat \Gamma^\mu.
\end{equation}
where the gauge source terms $\hat \Gamma^\mu(x^\alpha,g_{\alpha\beta})$ can have
functional dependence on the coordinates and the metric~\cite{Friedrich}. In
terms of the connection $\Gamma^\mu_{\alpha\beta}$, these harmonic conditions
take the form
\begin{equation}
   {\cal C}^\mu :=\Gamma^\mu -\hat \Gamma^\mu =0.
\end{equation}
where
\begin{equation}
     \Gamma^\mu = g^{\alpha\beta}\Gamma^\mu_{\alpha\beta}= 
              -\frac{1}{\sqrt {-g}}\partial_\alpha\gamma^{\alpha\mu}
\end{equation}
and $\gamma^{\mu\nu}=\sqrt{-g}g^{\mu\nu}$. 
The standard harmonic reduction of
the Einstein tensor (see e.g.~\cite{wald,Friedrich}) is 
\begin{equation}
     E^{\mu\nu}:= G^{\mu\nu} -\nabla^{(\mu}\Gamma^{\nu)} 
                 +\frac{1}{2}g^{\mu\nu}\nabla_\alpha \Gamma^\alpha,
		 \label{eq:e}
\end{equation}
where $\Gamma^\nu$ is treated formally as a vector in constructing the
``covariant'' derivatives $\nabla^{\mu}\Gamma^{\nu}$. Explicitly in terms
of the metric, 
\begin{eqnarray}
    2\sqrt{-g} E^{\mu\nu}&=&\partial_\alpha (g^{\alpha\beta}
           \partial_\beta \gamma^{\mu\nu})  
   -2\sqrt{-g}g^{\alpha\rho}g^{\beta\sigma}\Gamma^\mu_{\alpha\beta}
                   \Gamma^{\nu}_{\rho\sigma}
	-\sqrt{-g}(\partial_\alpha g^{\alpha\beta})\partial_\beta g^{\mu\nu}
   +\frac{1}{\sqrt{-g}}g^{\alpha\beta}(\partial_\beta g)\partial_\alpha g^{\mu\nu} 
             \nonumber \\
      &+&\frac{1}{2}g^{\mu\nu}\bigg (\frac{1}{2g\sqrt{-g}}g^{\alpha\beta}
                 (\partial_\alpha g)\partial_\beta g
        +\sqrt{-g}\Gamma^\rho_{\alpha\beta}\partial_\rho g^{\alpha\beta}
	+\frac{1}{\sqrt{-g}}(\partial_\beta g)\partial_\alpha g^{\alpha\beta}
	   \bigg ) .
\label{eq:efc}
\end{eqnarray}
Thus the standard harmonic evolution equations are
quasilinear wave equations  $E^{\mu\nu}=0$ for the components of the densitized
metric $\gamma^{\mu\nu}$. 
The principle part of (\ref{eq:efc}) has been expressed in the flux
conservative form
\begin{equation}
     \partial_\alpha (g^{\alpha\beta} \partial_\beta \gamma^{\mu\nu}).
\end{equation}      
In the particular case of the AwA gauge wave, the vanishing of the nonlinear
source terms then leads to the exact conservation laws for the quantitities
\begin{equation}
     Q^{\mu\nu}=-\int_V g^{t\alpha}\partial_\alpha\gamma^{\mu\nu} dV.
\end{equation}
The semi-discrete version of these conservation laws suppresses instabilities in
the AwA gauge wave test. (See~\cite{babev} for details.)

This standard treatment of the Cauchy problem in harmonic coordinates
generalizes in a straightforward way to include harmonic gauge source terms
and constraint adjustments of the form
\begin{equation}
         A^{\mu\nu}={\cal C}^\rho A^{\mu\nu}_\rho 
   (x^\alpha,g^{\alpha\beta},\partial_\gamma g^{\alpha\beta}). 
\end{equation} 
The reduced equations then become 
\begin{eqnarray}
   \tilde E^{\mu\nu} : &=&G^{\mu\nu} -\nabla^{(\mu} {\cal C}^{\nu)} 
                 +\frac{1}{2}g^{\mu\nu}\nabla_\alpha {\cal C}^\alpha
		 + A^{\mu\nu} \nonumber \\
                &=&E^{\mu\nu} +\nabla^{(\mu} \hat \Gamma^{\nu)} 
                 -\frac{1}{2}g^{\mu\nu}\nabla_\alpha \hat \Gamma^\alpha
		 + A^{\mu\nu}=0 
		 \label{eq:etilde}.
\end{eqnarray}
Since the gauge source terms and constraint adjustments do not enter the
principle part, (\ref{eq:etilde}) also constitute a system of quasilinear wave
equations with a well-posed Cauchy problem.

The solutions of the generalized harmonic evolution system (\ref{eq:etilde})
are solutions of the Einstein equations provided the harmonic constraints 
\begin{equation}
   {\cal C}^\mu :=\Gamma^\mu -\hat \Gamma^\mu =0
\end{equation}
are satisfied. The Bianchi identities, applied to (\ref{eq:etilde}), imply
that these constraints obey  the homogeneous wave equations
\begin{equation}
     \nabla^\alpha \nabla_\alpha {\cal C}^\mu +R^\mu_\nu {\cal C}^\nu 
                   -2\nabla_\nu ({\cal C}^\rho A^{\mu\nu}_\rho)=0,
     \label{eq:ahatconstr}
\end{equation}
where via (\ref{eq:etilde}) the Ricci tensor reduces to
\begin{equation}
     R^{\mu\nu} = \nabla^{(\mu}{\cal C}^{\nu)}- {\cal C}^\rho 
     (A^{\mu\nu}_\rho
     -\frac{1}{2}g^{\mu\nu}g_{\alpha\beta} A^{\alpha\beta}_{\rho}).
     \label{eq:aredricci}
\end{equation}
The well-posedness of the Cauchy problem for (\ref{eq:ahatconstr}) implies the
unique solution ${\cal C}^\mu =0$ in the domain of dependence of the initial
Cauchy hypersurface ${\cal S}$ provided the Cauchy data $\gamma^{\mu\nu}|_{\cal
S}$ and $\partial_t \gamma^{\mu\nu}|_{\cal S}$ satisfy
${\cal C}^\mu|_{\cal S}=\partial_t {\cal C}^\mu|_{\cal S}=0$
via (\ref{eq:etilde}). It is
straightforward to verify that Cauchy data on ${\cal S}$ which satisfy the
Hamiltonian and momentum constraints $G^t_\mu=0$ and the initial condition
${\cal C}^\mu=0$ also satisfy $\partial_t {\cal C}^\mu =0$ on ${\cal S}$ by
virtue of the reduced harmonic equations (\ref{eq:etilde}). In addition to the
standard Cauchy data for the $3+1$ decomposition, i.e. the intrinsic metric and
extrinsic curvature of ${\cal S}$ subject to the Hamiltonian and momentum
constraints, the only other free data are the initial choices of
$\gamma^{t\alpha}$, equivalent to the initial choices of lapse and shift.

The standard harmonic reduction (\ref{eq:e}) differs from Fock's~\cite{Fock}
harmonic formulation adopted in~\cite{harl} by the adjustment 
\begin{equation}
     A^{\mu\nu}=\frac{1}{2g}{\cal C}^{(\mu}g^{\nu)\alpha}\partial_\alpha g
          -\frac{1}{4g}g^{\mu\nu}  {\cal C}^\alpha \partial_\alpha g .
\label{eq:aold}
\end{equation}
In the tests of the boundary algorithms in Sec.~\ref{sec:tests}, we have
examined the effects of the adjustments
\begin{eqnarray}
     A^{\mu\nu} &=& -\frac {a_1}{\sqrt{-g}} {\cal C}^\alpha 
              \partial_\alpha (\sqrt{-g}g^{\mu\nu}) 
	      \label{eq:a1adj} \\
    A^{\mu\nu} &=& \frac{a_2 {\cal C}^\alpha \nabla_\alpha t}
      {e_{\rho\sigma}{\cal C}^\rho {\cal C}^\sigma}
         {\cal C}^{\mu} {\cal C}^{\nu} 
	 \label{eq:a2adj} \\ 
 A^{\mu\nu} &=& -\frac {a_3}{\sqrt{-g^{tt}}} {\cal C}^{(\mu}\nabla^{\nu)}t ,     
\label{eq:a3adj}
\end{eqnarray}
where the $a_i>0$ are positive adjustable parameters and 
\begin{equation}
        e_{\rho\sigma}=g_{\rho\sigma}- 
          \frac{2}{g^{tt}}(\nabla_\rho t)\nabla_\sigma t
\end{equation}
is the natural metric of signature $(++++)$ associated with the Cauchy slicing.
The adjustments (\ref{eq:a1adj}) and (\ref{eq:a2adj}) were effective in
suppressing instabilities in the shifted gauge wave tests without
boundaries~\cite{babev}. (For numerical purposes, the limit as $C^\mu
\rightarrow 0$ in (\ref{eq:a2adj}) can be regularized by a small positive
addition to the denominator.) The adjustment (\ref{eq:a3adj}) leads to
constraint damping~\cite{constrdamp} in the linear regime and has been used
effectively in black hole simulations~\cite{pret2} but was not effective in the
nonlinear shifted gauge wave test.

\section{Well-posedness of the homogeneous initial-boundary value problem}
\label{sec:hom}

In prior work~\cite{harl} we showed how the well-posedness of the Cauchy
problem for the standard harmonic formulation of Einstein's equations extends
to the IBVP with homogeneous boundary data. Here we show how the well-posedness
of the IBVP extends to include harmonic gauge forcing terms and constraint
adjustments. Our approach is based upon a theorem of Secchi~\cite{secchi2} for
first order, quasilinear, symmetric hyperbolic systems.  For that purpose, we
recast the reduced system (\ref{eq:etilde}) in first order form in terms of the
50-dimensional column vector ${\bf u}={}^T(\gamma^{\alpha\beta},{\cal
T}^{\alpha\beta}, {\cal X}^{\alpha\beta}, {\cal Y}^{\alpha\beta},{\cal
Z}^{\alpha\beta})$ (the transpose of
the row vector ${}^T {\bf u}$) where ${\cal T}^{\alpha\beta}= \partial_t
\gamma^{\alpha\beta}$, ${\cal X}^{\alpha\beta}=\partial_x
\gamma^{\alpha\beta}$, ${\cal Y}^{\alpha\beta}= \partial_y
\gamma^{\alpha\beta}$ and ${\cal Z}^{\alpha\beta}=\partial_z
\gamma^{\alpha\beta}$.
Equation (\ref{eq:etilde}) then has the quasilinear,
symmetric hyperbolic form 
\begin{equation} 
      {\bf A}^t \partial_t {\bf u}+{\bf A}^i\partial_i {\bf u} = {\bf S} \label{eq:sev} 
\end{equation} 
where ${\bf A}^t$ and ${\bf A}^i$ are 50-dimensional symmetric matrices, with
${\bf A}^t$ positive definite, and ${\bf S}$ is a 50-dimensional column vector,
all depending only on the components of ${\bf u}$. For explicit values of the
matrices ${\bf A}^t$ and ${\bf A}^i$ see App.~\ref{app:scalar}.

Consider the IBVP for the reduced Einstein equations for the generalized
harmonic system in the domain $t\ge 0$, $x\le 0$, with timelike boundary ${\cal
B}$ at $x=0$. (Any boundary can be constructed by patching together such
pieces.) Secchi's theorem establishes well-posedness of the IBVP for
(\ref{eq:sev}) under weak regularity assumptions by imposing a
homogeneous boundary condition based upon the energy norm 
\begin{equation} 
      E=\frac{1}{2}\int_ {}^T{\bf u u}dxdydz.
\end{equation} 
and the associated local energy flux across the boundary
\begin{equation} 
      {\cal F}^x=\frac{1}{2} {}^T{\bf u}{\bf A}^x {\bf u}.
\end{equation}  

The chief requirement for a
well-posed IBVP is that the boundary condition be maximally dissipative, which
allows energy estimates to be established.  Explicitly, the
requirements at $x=0$ are 
\begin{itemize}

\item that the kernel of the boundary matrix ${\bf A}^x$ must have
constant dimension,

\item  that the boundary condition must take the matrix form ${\bf Mu}=0$ for
homogeneous boundary data, where ${\bf M}$ is independent of ${\bf u}$ and has
maximal dimension consistent with the system of equations

\item and, for all ${\bf u}$ satisfying the boundary condition, that the
 boundary flux satisfy the inequality 
\begin{equation}
     {\cal F}^x \ge 0.
\end{equation}

\end{itemize}
In addition, compatibility conditions 
between the Cauchy and boundary data at $t=x=0$ must be satisfied.

As a result of reducing the harmonic Einstein equations to first order
symmetric hyperbolic form, there are variables associated with characteristics
which propagate tangential to the boundary. Secchi's theorem applies to this
case of a {\em characteristic}  boundary in which the matrix ${\bf A}^x$ is
degenerate. The matrix ${\bf M}$ can depend explicitly on the coordinates but
not upon the evolution variables ${\bf u}$. The maximal dimension of ${\bf M}$
is directly related to how many variables propagate toward the boundary from
the exterior. Variables in the kernel of ${\bf A}^x$, which propagate
tangential to the boundary, and variables which propagate from the interior
toward the boundary require no boundary condition. The possible choices of
matrix ${\bf M}$ correspond to the nature of the boundary condition, e.g.
Dirichlet, Sommerfeld or Neumann. 

We now formulate such maximally dissipative boundary conditions for the
reduced harmonic system. The principal part consists of the term
$g^{\alpha\beta}\partial_\alpha\partial_\beta \gamma^{\mu\nu}$, which
represents a wave operator acting on the 10 components $\gamma^{\mu\nu}$.
As a result, the energy flux ${\cal F}^x$ equals the sum of 10
individual fluxes, each of which corresponds to the standard energy flux
constructed from the stress-energy tensor obtained by treating each
component of $\gamma^{\mu\nu}$ as a massless scalar field. Thus we can
pattern our analysis of the IBVP on the scalar wave equation
\begin{equation}
         g^{\mu\nu}\partial_\mu \partial_\nu \Phi = 0,
	 \label{eq:sweq}
\end{equation}
which is discussed in App.~\ref{app:scalar}.  
When $\Phi$ is regarded as an independent
scalar field rather than as a component of $\gamma^{\mu\nu}$, the linearity
of (\ref{eq:sweq}) implies 
well-posedness for any of the dissipative boundary conditions 
discussed in App.~\ref{app:scalar}, e.g. generalized Dirichlet, Neumann or
Sommerfeld boundary conditions. 

We can decompose the boundary conditions for the full gravitational problem
into the five 10-dimensional subspaces corresponding to $\gamma^{\alpha\beta}$,
${\cal T}^{\alpha\beta}$, ${\cal X}^{\alpha\beta}$, ${\cal Y}^{\alpha\beta}$
and ${\cal Z}^{\alpha\beta}$. The subspace corresponding to
$\gamma^{\alpha\beta}$ lies in the kernel of ${\bf A}^x$ and requires no
boundary condition, i.e the boundary values of $\gamma^{\alpha\beta}$ are
determined by boundary values of the other variables. The remaining four
subspaces require boundary conditions analogous to the scalar field case.  The
extra complications in the quasi-linear gravitational case are to check that
the constraints are satisfied and that the linear subspace ${\bf Mu}=0$ is
specified independent of ${\bf u}$. 

For that purpose, we adapt the harmonic coordinate system to the boundary. In
doing so, we use the $3+1$ decomposition natural to the boundary at $x=0$ and
write  $x^\alpha=(x^a,x)$, where $x^a=(t,y,z)$. Our approach is motivated by
the observation that the initial value problem is well-posed so that reflection
symmetric Cauchy data in the neighborhood $-\epsilon<x<\epsilon$ produces a
solution with even parity. From the point of view of the IBVP in the region
$x\le 0$, this solution induces homogeneous boundary data of either the Neumann
or Dirichlet type on each metric component, depending upon how that component
transforms under reflection.

\subsection{The boundary gauge}
\label{sec:bgauge}

Local to the boundary ${\cal B}$, it is always possible to choose Gaussian
normal coordinates  $(\tilde x^a,\tilde x)$ so that the boundary is given
by $\tilde x =0$ and $g^{\tilde x \tilde a}=0$ vanishes at the boundary.
Now consider a coordinate transformation to generalized harmonic
coordinates  $(x^a,x)$ satisfying $\Box x^\alpha= -\hat \Gamma^\alpha$. Since
$g^{\tilde x \tilde a}=0$, the analysis in App.~\ref{app:scalar} shows
that it is possible to solve this wave equation with either the
homogeneous Dirichlet condition $\Phi=0$ or the homogeneous Neumann
condition $\partial_x \Phi =0$. For the harmonic coordinate $x$, we choose
the Dirichlet condition $x|_{\cal B}=0$, so that ${\cal B}$ remains
located at $x=0$; and for the remaining harmonic coordinates, we choose
the Neumann condition  $\partial_{\tilde x} x^a|_{\cal B}=0$. 

These choices exhaust the boundary freedom in constructing generalized
harmonic coordinates. After transforming to these harmonic coordinates,
the metric satisfies 
\begin{equation}
     \gamma^{xa}|_{\cal B}=0.
\label{eq:bgauge}     
\end{equation}
We enforce this as the dissipative Dirichlet boundary condition
\begin{equation}
     {\cal T}^{xa}|_{\cal B}=0,
\label{eq:homdir}     
\end{equation}
subject to initial data satisfying (\ref{eq:bgauge}).

In this boundary gauge, the results of App.~\ref{app:scalar} imply that
the homogeneous Neumann boundary conditions
\begin{equation}
     {\cal X}^{xx}|_{\cal B}={\cal X}^{ab}|_{\cal B}=0,
\label{eq:homneum}     
\end{equation} 
are also dissipative, as well as the Sommerfeld-like
conditions
\begin{equation}
   ({\cal T}^{xx} + {\cal X}^{xx})|_{\cal B}=
     ({\cal T}^{ab} + {\cal X}^{ab})|_{\cal B}=0.
\label{eq:homsom}     
\end{equation} 

Given that the boundary gauge condition (\ref{eq:bgauge})  is satisfied, any
combination of Dirichlet, Sommerfeld-like and Neumann boundary conditions on
the remaining components leads to a well-posed IBVP for the reduced
generalized harmonic system (\ref{eq:etilde}). Note, however, that a Sommerfeld
condition corresponding to the null direction defined by normal to the
boundary, 
 \begin{equation}
   ({\cal T}^{xx} + \frac{\sqrt{-g_{tt}}}
        {\sqrt{g_{xx}}}{\cal X}^{xx})|_{\cal B}=
    ({\cal T}^{ab} 
       +\frac{\sqrt{-g_{tt}}}{\sqrt{g_{xx}}}{\cal X}^{ab})|_{\cal B}=0,
\label{eq:homgsom}     
\end{equation} 
does not satisfy the technical condition in Secchi's theorem
that $M$ be independent of $u$.

\subsection{The homogeneous constraint-preserving boundary condition}
\label{sec:satcon}

The constraints ${\cal C}^\mu=\Gamma^\mu-\hat \Gamma^\mu$ satisfy the
homogeneous
wave equation (\ref{eq:ahatconstr}), which can be reduced to symmetric
hyperbolic form. Consequently, we can force them to
vanish by choosing boundary conditions for the reduced system which force
${\cal C}^\mu$ to satisfy a maximally dissipative homogeneous boundary
condition. In doing so, we shall require that the normal components
of the gauge source function and constraint adjustment 
vanish on the boundary, i.e.
\begin{eqnarray}
      \hat \Gamma^x|_{\cal B}&=&0 \nonumber \\
             A^{ax}|_{\cal B}&=&0 .
\label{eq:hathzboun}
\end{eqnarray}
These requirements are a consequence of the reflection properties of the
boundary in the homogeneous case.

We proceed by imposing boundary conditions on the evolution system consisting
of the Dirichlet boundary-gauge conditions (\ref{eq:bgauge}) and the Neumann
conditions (\ref{eq:homneum}), which imply
\begin{equation}
    \gamma^{ax}|_{\cal B} = \partial_x \gamma^{xx}|_{\cal B}=
         \partial_x \gamma^{ab}|_{\cal B} = \partial_x g|_{\cal B}=0.
\label{eq:bc}
\end{equation}
As we shall show, (\ref{eq:bgauge}) and (\ref{eq:bc}) further
imply that the constraints obey the maximally dissipative homogeneous boundary
conditions
\begin{eqnarray}
     \partial_t {\cal C}^x|_{\cal B} &=& 0
\label{eq:hbmdz}   \\
   \partial_x {\cal C}^a|_{\cal B} &=& 0.
\label{eq:hbmda}     
\end{eqnarray}

Note that (\ref{eq:hbmdz}) and (\ref{eq:hbmda}) are satisfied by any
geometry which has local reflection symmetry with respect to the boundary if
the gauge source functions $\hat \Gamma^\alpha$ share this symmetry (as a
vector). In order to verify (\ref{eq:hbmdz}), we note that
\begin{equation}
      \partial_t{\cal C}^x = -\partial_t \bigg (\frac{1}{\sqrt{-g}}
      (\partial_x \gamma^{xx} + \partial_a \gamma^{xa}) +\hat \Gamma^x \bigg )
\end{equation}
vanishes on the boundary as a consequence of (\ref{eq:hathzboun}) and
(\ref{eq:bc}).

Similarly, in order to verify (\ref{eq:hbmda}), we note that
\begin{equation}
    \partial_x {\cal C}^a= -\partial_x \bigg( \frac{1}{\sqrt{-g}}
        (\partial_x\gamma^{ax} +\partial_b \gamma^{ab})+ \hat \Gamma^a \bigg )
\end{equation}
reduces on the boundary to
\begin{equation}
    \partial_x {\cal C}^a|_{\cal B}=
        -\bigg (\frac{1}{\sqrt{-g}}\partial_x^2 \gamma^{ax} +
                      \partial_x \hat \Gamma ^a \bigg )|_{\cal B}.
\label{eq:caz}
\end{equation}
Now we use the $\tilde E^{ax}=0$ component of the evolution equation to
eliminate $\partial_x^2 \gamma^{ax}$. Substitution of the assumed boundary
conditions into (\ref{eq:efc}) gives
\begin{equation}
     E^{ax}|_{\cal B} = \frac{g^{xx}}{2\sqrt{-g}}
         \partial_x^2 \gamma^{ax}|_{\cal B}.
\end{equation}
Then (\ref{eq:etilde}) gives 
\begin{equation}
           0=\tilde E^{ax}|_{\cal B} = 
           \frac{g^{xx}}{2}\bigg(\frac{1}{\sqrt{-g}}\partial_x^2 \gamma^{ax}
        + \partial_x \hat \Gamma^a \bigg )|_{\cal B}
          +A^{ax}|_{\cal B},
\end{equation}
which, together with (\ref{eq:hathzboun}) and (\ref{eq:caz}), establishes
(\ref{eq:hbmda}). 

In summary, the maximally dissipative boundary conditions  (\ref{eq:bgauge})
and (\ref{eq:homneum}) for the evolution variables ${\bf u}$ imply the
maximally dissipative homogeneous boundary conditions (\ref{eq:hbmdz}) and
(\ref{eq:hbmda}) for the constraints. This establishes the boundary data
necessary to show that the harmonic constraints ${\cal C}^\mu=0$ propagate if
the initial Cauchy data satisfy the Hamiltonian and momentum constraints.

The remaining ingredient necessary for a well-posed initial-boundary problem is
the consistency between the initial Cauchy data and the boundary data. The
order of consistency determines the order of differentiability of the solution.
The simplest case to analyze is prescription of initial data with locally
smooth reflection symmetry at the boundary, e.g Cauchy data
$\gamma^{ab}(0,x,y,z)$ for $x\le 0$ whose extension to $x\ge 0$ by reflection
would yield a  $C^\infty$ function in the neighborhood of $x=0$. Such data are
automatically consistent with the homogeneous boundary conditions given above. 

We could also have established the boundary conditions (\ref{eq:hbmda}) for the
constraints indirectly, but with a more geometric argument, by noting that the
boundary conditions (\ref{eq:bc}) for the metric imply that the extrinsic
curvature of the boundary vanishes, which in turn implies that the $G^{ax}$
components of the Einstein tensor vanish at the boundary (see 
App.~\ref{app:extrinsic}). Thus, along with the evolution equation $\tilde
E^{ax}=0$,  (\ref{eq:etilde}) implies
\begin{equation}
    -\nabla^{(a} {\cal C}^{x)} 
                 +\frac{1}{2}g^{ax}\nabla_\alpha {\cal C}^\alpha
		 + A^{ax} =0 .
\end{equation} 
This, combined with (\ref{eq:hathzboun}) - (\ref{eq:hbmdz}), now gives
(\ref{eq:hbmda}). We take this geometrical approach in the next section in
analyzing the difficulties underlying extending these results to include
inhomogeneous boundary data. 

\section{Inhomogeneous boundary data}
\label{sec:inhom}

In the preceding section we showed that the generalized harmonic evolution
system $\tilde E^{\mu\nu}=0$ gives rise to a well-posed constraint-preserving
IBVP with a variety of maximally dissipative homogeneous boundary conditions.
The well-posedness of this IBVP for the reduced system extends by standard
arguments to include inhomogeneous data. However, only for special choices of
the boundary data is the solution guaranteed to satisfy the constraints. For
example, in the simulation of a known analytic solution to Einstein's
equations, when the initial data and boundary data are supplied by their
analytic values, the well-posedness of the IBVP guarantees a unique and
therefore constraint-preserving solution. This example is important in code
development for carrying out tests of the type reported in
Sec.~\ref{sec:tests}. Similarly, when an exterior solution of the Einstein
equations is supplied in another patch overlapping the boundary, the induced
boundary data are constraint-preserving. This possibility arises in
Cauchy-characteristic matching. Except for such examples, the specification of
free boundary data for the reduced system does not in general lead to a
solution of the Einstein equations.
 
We now formulate conditions on the inhomogeneous
boundary data which are sufficient to
guarantee that the IBVP for the generalized harmonic system leads to a solution
that satisfies the constraints. However, because the matrix ${\bf M}$ used in
formulating these constraint-preserving boundary conditions depends on the
evolution variables ${\bf u}$, the well-posedness of the resulting IBVP follows
from Secchi's theorems only for ``small'' boundary data, in the sense of
linearization about a solution with homogeneous data. Well-posedness for
``large'' boundary data remains an open question.

As in the homogeneous case, we consider the IBVP in the domain $t\ge 0$, $x\le
0$.  We first generalize the boundary gauge condition (\ref{eq:bgauge}). Given
harmonic coordinates $x^\nu$ such that $\gamma^{xa}|_{\cal B}=0$ at $x=0$,
consider the effect of a harmonic coordinate transformation $\tilde
x^\mu(x^\nu)$ satisfying $\Box \tilde x^\mu=-\hat \Gamma^\mu$.  For simplicity,
we assign initial Cauchy data $\tilde x^\mu = x^\mu$ and $\partial_t \tilde
x^\mu =\partial_t x^\mu $ at $t=0$ and require that the boundary remain located
at $\tilde x=0$. By uniqueness of solutions to the IBVP for the wave equation,
this Dirichlet boundary data for $\tilde  x$ determines that $\tilde x=x$
everywhere in the evolution domain so that $\tilde x^\mu =(\tilde x^a, x)$. The
coordinates $\tilde x^a$ are uniquely determined by the Neumann boundary data
$q^a=\partial_x \tilde x^a|_{\cal B}$, which can be assigned freely. Under this
transformation,
$\gamma^{\tilde x \tilde a}=q^a \gamma^{\tilde x \tilde x}$ at the boundary, so that
$\gamma^{\tilde x \tilde a}/\gamma^{\tilde x\tilde x}$
can be freely specified as boundary data $q^a (x^b)$. (Consistency conditions
would require that $q^a=0$ at $t=x=0$. However, the initial value of $q^a$ can
be freed up by a deformation of the initial Cauchy hypersurface in the
neighborhood of the boundary.)  In the remainder of this section, we assume
that this transformation has been made and that the $\tilde x^\mu(x^\nu)$
coordinates have been relabeled $x^\mu$. Thus, in this coordinate system,
\begin{equation}
          q^a(x^b)=\frac{\gamma^{xa}}{\gamma^{xx}}|_{\cal B}
\label{eq:qa}
\end{equation}
is free boundary data representing the choice of boundary gauge at $x=0$.

Although it might seem that such a gauge transformation would not disturb the
well-posed nature of the IBVP problem for the reduced system, already
(\ref{eq:qa}) introduces the complication that Dirichlet boundary data for
$\gamma^{xa}$ cannot be determined from $q^a$ unless $\gamma^{xx}$ is known
on the boundary. But the boundary condition (\ref{eq:bc}) for $\gamma^{xx}$
is of Neumann type so that $\gamma^{xx}|_{\cal B}$ can only be determined
after carrying out the evolution and, in particular, depends on the initial
Cauchy data. This is the first source of the ${\bf u}$-dependence of the
matrix ${\bf M}$ expressing the boundary condition. It would seem that this
problem might be avoided by using $\gamma^{xa}/\gamma^{xx}$ as an evolution
variable but, as will become apparent, more complicated (differential)
problems of this type arise in introducing inhomogeneous constraint
preserving boundary data for the remaining variables.

The complete set of free boundary data for the generalized harmonic system,
which in the inhomogeneous case replaces the maximally dissipative homogeneous
boundary conditions (\ref{eq:bgauge}) and (\ref{eq:bc}), consist of 
\begin{equation}
   {\bf q}=(q^a,q^{xx},q^{ab})=(q^a,q^\mu\partial_\mu \gamma^{xx},
        q^\mu\partial_\mu \gamma^{ab})|_{\cal B},
	\label{eq:qdata}
\end{equation}
 where, in accord with (\ref{eq:qa}),
\begin{equation}
    q^\mu =(q^a,1) = \frac {1}{g^{xx}} \nabla^\mu x 
        = \frac {1}{\sqrt{g^{xx}}}  n^\mu ,
\label{eq:qalpha}
\end{equation}
with $n^\mu$ the unit outward normal to the boundary. 

Note that a dissipative Neumann condition must be formulated in terms of the
normal derivative $q^{\mu}\partial_\mu$ and not the $\partial_x$ derivative
used in the previous section where $q^a=0$. For this purpose, it is useful to
introduce the projection tensor into the tangent space of the boundary,
\begin{equation}
    h_{\mu}^{\nu} =\delta_{\mu}^{\nu}-n_\mu n^\nu.
\end{equation}
The extrinsic curvature tensor of the boundary is given by
\begin{equation}
     K^{\mu\nu}=h^\mu_\alpha h^\nu_\beta
           \nabla^\alpha n^\beta.
\label{eq:kmunu}	   
\end{equation}
Its explicit expression in terms of the harmonic variables used here is given
in App.~\ref{app:extrinsic}.

The inhomogeneous boundary data ${\bf q}$ must be constrained if the Einstein
equations are to be satisfied.  The subsidiary system of equations which
governs the constraints requires a slightly more complicated treatment than in
the previous section. For its analysis, we represent the constraints by ${\bf
C}=(C^x,C_a=g_{a\mu}C^\mu)$. Since $C^\mu =(C^x n_x -C_a n^a)n^\mu +C_a
g^{a\mu}$, it is clear that ${\bf C} =0$ is equivalent to $C^\mu =0$. It is
straightforward to show, beginning with (\ref{eq:ahatconstr}), that ${\bf C}$
satisfies a wave equation of the homogeneous form 
\begin{eqnarray}
    g^{\mu\nu}\partial_\mu \partial_\nu {\bf C}
      + {\bf P}^\mu \partial_\mu {\bf C} + {\bf Q}{\bf C} =0,
\label{eq:bhwave}
\end{eqnarray}
where the elements of the square  matrices ${\bf P}^\mu$ and ${\bf Q}$ depend
algebraically on the evolution variables ${\bf u}$. This system can be
converted to symmetric  hyperbolic first order form. Uniqueness of the solution
to the IBVP then ensures that the generalized harmonic constraints are
satisfied in the evolution domain provided the initial Cauchy data satisfy
$C^\mu=\partial_t C^\mu=0$  and that ${\bf C}$ satisfies a maximally
dissipative homogeneous boundary condition.

\subsection{The boundary system}

We choose as the maximally dissipative homogeneous boundary condition for ${\bf
C}$
\begin{eqnarray}
     \sqrt{g^{xx}} n_\mu C^\mu|_{\cal B} =C^x|_{\cal B} &=& 0 
\label{eq:gbmdz}   \\
 \frac {1}{\sqrt{g^{xx}}}h_a^\nu n^\mu \partial_\mu C_\nu|_{\cal B}= 
         q^\mu \partial_\mu C_a|_{\cal B} &=& 0.
\label{eq:gbmda}     
\end{eqnarray}
Referring to (\ref{eq:sflux}), the resulting boundary flux associated with each
component of ${\bf C}$ satisfies the dissipative condition ${\cal F}^x=0$.
These boundary conditions, along with the constraints on the initial Cauchy
data, guarantee that solutions of the reduced system satisfy Einstein's
equations. For simplicity in working out their consequence, we require the
gauge source terms to satisfy
\begin{equation}
     n_\alpha \hat \Gamma^\alpha|_{\cal B}=0,
\label{eq:athzboun}
\end{equation}
as in (\ref{eq:hathzboun}) for the homogeneous case, but
we make no assumption about the behavior of the constraint adjustment at
the boundary. 

The boundary condition (\ref{eq:gbmdz}), together with (\ref{eq:athzboun}),
then implies $\Gamma^x=\hat \Gamma^x=0$ so that $\partial_a \gamma^{xa} +
\partial_x \gamma^{xx}=0$. Using (\ref{eq:qa}) and (\ref{eq:qdata}), this
reduces to
\begin{equation}
   q^{xx}=-\gamma^{xx}|_{\cal B}\partial_a q^a, 
\label{eq:gqzz}
\end{equation} 
which provides the constrained Neumann boundary data $q^{xx}$.

The boundary condition (\ref{eq:gbmda}) contains the term $\partial_x^2
\gamma^{xa}$, which must be eliminated by using the reduced equations. We
proceed by using
(\ref{eq:gbmdz}) to obtain, at the boundary, 
\begin{eqnarray}
      h_\mu^\nu n_\beta (\nabla_\nu C^\beta +  \nabla^\beta C_\nu)
     &=& -h_\mu^\nu C^\beta \nabla_\nu n_\beta
             +h_\mu^\nu n^\beta \nabla_\beta C_\nu  \\
      &=& -K_{\mu\beta} C^\beta 
              +h_\mu^\nu{\cal L}_n C_\nu
	      - h_\mu^\nu C_\beta \nabla_\nu n^\beta \\
	    &=& -2K_{\mu\beta} C^\beta 
              +h_\mu^\nu {\cal L}_n C_\nu ,
\end{eqnarray}
where $K^{\mu\nu}$ is the extrinsic curvature introduced in
(\ref{eq:kmunu}). In the $x^a$-coordinates of the boundary, this implies
\begin{equation}
     (\nabla_a C^x +  \nabla^x C_a)=
           -2\sqrt{g^{xx}}K_{ab}C^b
      +g^{xx}q^\mu \partial_\mu C_a+g^{xx}C_b \partial_a q^b.
\label{eq:baz}
\end{equation}
Assuming that the reduced equations (\ref{eq:etilde}) are satisfied,
the boundary constraint (\ref {eq:gbmda}) can then be re-expressed as
\begin{equation}
   h_a^\alpha n_\beta G_\alpha^\beta = \frac{1}{\sqrt{g^{xx}}}G_a^x
   = -K_{ab}C^b +\frac{1}{2}\sqrt{g^{xx}}C_b \partial_a q^b 
          -\frac{1}{\sqrt{g^{xx}}}A_a^x . 
\end{equation} 
which contains no second $x$-derivatives of the metric. 
Here $h_a^\alpha n_\beta G_\alpha^\beta$ is the ``boundary momentum''
(\ref{eq:bmom}) so that this equation can be re-expressed as
\begin{equation}
  \bigg ( D_b (K_a^b-\delta_a^b K) +K_{ab} C^b
        -\frac{\sqrt{g^{xx}}}{2} C_b \partial_a q^b
        +\frac{1}{\sqrt{g^{xx}}}A_a^x  \bigg )|_{\cal B}=0. 
\label{eq:gakconstr}
\end{equation} 
This is a system of  equations intrinsic to the boundary which provides the
means for prescribing constraint-preserving values for the boundary data
$q^{ab}$. 

\subsection{Constraint-preserving boundary data}

The boundary system (\ref{eq:gakconstr}) can be recast as a symmetric
hyperbolic system which determines constrained values of the Neumann data
$q^{ab}$ in terms of the free (boundary gauge) data $q^a$ and the boundary
values of the other evolution variables  ${\cal
U}:=[\gamma^{ab},\gamma^{xx},\partial_x \gamma^{xa})]_{\cal B}$ which cannot be
freely specified. In doing so, there is considerable algebra in expressing
(\ref{eq:gakconstr}) in terms of the evolution variables and boundary data. 

First, the covariant derivative has to be explicitly worked out, as in
(\ref{eq:scd}) where we now set $S^{ab}=K^{ab}-h^{ab}K$. From
(\ref{eq:gakconstr}), we then find (on the boundary)
\begin{equation}
    \partial_b(\sqrt{-h_{\cal B}}S^{ab}) = P^a ,
\label{eq:gakconstr5}
\end{equation}
where
\begin{equation}
   P^a =-\sqrt{-h_{\cal B}}(h^{ac}S^{bd}
        -\frac{1}{2}h^{ad}S^{bc})\partial_d h_{bc}
          	  -\sqrt{-h_{\cal B}}K^{ab} C_b 
           +\frac{1}{2}\sqrt{-g^{xx}h_{\cal B}} h^{ab} C_d\partial_b q^d 
           -\sqrt{\frac{-h_{\cal B}}{g^{xx}}} h^{ab} A_b^x.
\end{equation} 

Here
\begin{equation}
   h^{ab} A_b^x = A^{ax} - q^a A^{xx}.
\end{equation}
For the adjustment  (\ref{eq:a1adj}), 
\begin{equation}
     h^{ab} A_b^x  =-a_1 g^{xx}C^b\partial_b q^a.
\end{equation}
For the adjustment (\ref{eq:a2adj}), $h^{ab} A_b^x =0$.
For the adjustment (\ref{eq:a3adj}),
\begin{equation}
     h^{ab} A_b^x  =-\frac {a_3 g^{xx}}{2 \sqrt{-g^{tt}}}C^a q^t.
\end{equation}
For Fock's adjustment (\ref{eq:aold}), 
\begin{equation}
     h^{ab} A_b^x  =\frac {1}{4g} g^{xx}  C^a q^\mu \partial_\mu g.
\end{equation}

From (\ref{eq:kmt}), the individual terms
constituting $S^{ab}$ are
\begin{eqnarray}
  S^{ab} &=& \frac{\sqrt{g^{xx}}}{2}(h^{ac}\partial_c q^b
               + h^{bc}\partial_c q^a  -2 h^{ab}\partial_c q^c) 
	       \nonumber \\ 
        	&-& \frac{g^{xx}}{2\sqrt{-h_{\cal B}}}\big (q^{ab} 
		 +(\frac {h^{ab}}{g^{xx}}+q^a q^b)q^{xx}
	   -q^a q^\lambda \partial_\lambda \gamma^{xb} 
	   -q^b q^\lambda \partial_\lambda \gamma^{xa} \big ),
\end{eqnarray}
where it is assumed that $q^{xx}$ is replaced by its constrained value
(\ref{eq:gqzz}). 
We write
\begin{equation}
    \sqrt{-h_{\cal B}} S^{ab}=\frac{g^{xx}}{2} (-q^{ab} +T^{ab}),
\end{equation}
where
\begin{eqnarray}
  T^{ab} &=& \sqrt{-g}(h^{ac}\partial_c q^b
               + h^{bc}\partial_c q^a  -2h^{ab}\partial_c q^c) 
	       \nonumber \\ 
        	&-&(\frac {h^{ab}}{g^{xx}}+q^a q^b)q^{xx}
	   +q^a q^\lambda \partial_\lambda \gamma^{xb} 
	   +q^b q^\lambda \partial_\lambda \gamma^{xa} 
\end{eqnarray}
does not contain $q^{ab}$. Substitution of these pieces into
(\ref{eq:gakconstr5}) gives
\begin{equation}
    \partial_b q^ {ab} = V^a  ,
    \label{eq:qab}
\end{equation}
where
\begin{equation}
    V^a = \frac{1}{g^{xx}}\bigg (-q^ {ab}\partial_b g^{xx}
            + \partial_b (g^{xx}T^ {ab})-2P^a \bigg ).
\end{equation}
This  system of partial differential equations, with principle part $\partial_b
q^ {ab}$, governs the boundary data $q^{ab}$ in terms of the quantities ${\cal
U}$ and $q^a$.  Here $C^a$ (which appears in $P^a$) must be expressed in terms
of evolution variables according to 
\begin{equation}
    C^a=-\frac{1}{\sqrt{-g}}(\partial_b \gamma^{ab}+\partial_x \gamma^{ax}) 
                              -\hat \Gamma^a.
\end{equation}

We formulate a symmetric hyperbolic system
by setting 
\begin{equation} 
    q^{ab}= {\cal Q}^{ab} - \frac{1}{2} \eta^{ab} \eta_{cd} {\cal Q}^{cd}, 
\label{eq:qfromy}    
  \end{equation}
where $\eta_{ab}$ is the auxiliary Minkowski metric associated with the
coordinates of the boundary.
Equation (\ref{eq:qab}) then becomes
\begin{equation}
   \partial_b {\cal Q}^{ab}
   - \frac{1}{2} \eta^{ab} \partial_b \left( \eta_{cd}  {\cal Q}^{cd} \right) 
     = V^a,
\label{eq:Yeq2}
\end{equation}
Next we introduce the ``2+1'' decomposition of the boundary $x^a=(t,x^A)
=(t,y,z)$ and the quantities $\phi = \frac{1}{2} {\cal Q}^{tt}$ and ${\cal Q}^A
={\cal Q}^{tA}$. Then (\ref{eq:Yeq2}) gives the symmetric hyperbolic system
\begin{eqnarray}
  \partial_t \phi + \partial_B {\cal Q}^{B} 
   &=& V^t 
   - \frac{1}{2} \partial_t \left[ \delta_{CD} {\cal Q}^{CD} \right]
    \nonumber \\
   \partial_t {\cal Q}^{A} 
   + \delta^{AB} \partial_B  \phi
    &=& V^A 
     + \frac{1}{2} \delta^{AB} \partial_B \left[ \delta_{CD} {\cal Q}^{CD} \right] 
     - \partial_B {\cal Q}^{AB}
\label{eq:Yeq3}
\end{eqnarray}
for the variables $\phi$ and ${\cal Q}^A$. Assuming that the boundary values of
${\cal U}$ are known, (\ref{eq:Yeq3}) determines the constrained boundary
values of $\phi$ and ${\cal Q}^A$ in terms of the free boundary data
$(q^a,{\cal Q}^{AB})$. 

Constraint-preserving boundary data ${\bf q}$ for the reduced system is determined
from the free data $(q^a,{\cal Q}^{AB})$, with $q^{xx}$ obtained
from (\ref{eq:gqzz}) and $q^{ab}$ obtained from (\ref{eq:qfromy}) and
(\ref{eq:Yeq3}). This data ${\bf q}$ leads to the  maximally dissipative
homogeneous boundary data (\ref{eq:gbmdz}) and (\ref{eq:gbmda}) for the
subsidiary system (\ref{eq:bhwave}) which governs the constraints.
Consequently, {\em any solution of the IBVP for the reduced equations with this
boundary data is also a solution of the Einstein equations} (assuming that the
initial Cauchy data satisfies the constraints).

In the case of the  homogeneous boundary data ${\bf q}=0$ treated in
Sec.~\ref{sec:hom},  (\ref{eq:gqzz}) and (\ref{eq:gakconstr}) are automatically
satisfied 
and the IBVP is well-posed. The appearance of the  quantities ${\cal U}$ in
(\ref{eq:gqzz}) and (\ref{eq:gakconstr}) prevent application of Secchi's
theorem to prove the well-posedness of the inhomogeneous IBVP. However,
these theorems do imply a well-posed IBVP for boundary data $\delta {\bf
q}$ linearized off a solution with homogeneous data, which provides background
values of ${\cal U}$. From the analytical point  of view, such linearized
results often provide the first step in an iterative argument to show that the
full system is well-posed. From a numerical point of view, which is our prime
interest here, it suggests that the evolution code should be structured so that
the values of ${\cal U}$ are updated before updating the values of ${\bf q}$ 
in an attempt to force ${\cal U}$ into such a background role.

\section{Finite difference implementation}
\label{sec:imp}

A first differential order formalism is useful for applying the theory of
symmetric hyperbolic systems but in a numerical code it would introduce extra
variables and the associated nonphysical constraints. For this reason, we base
our code on the natural second order form of the quasilinear wave equations
which comprise the reduced harmonic system (\ref{eq:etilde}). They are finite
differenced in the flux conservative form 
\begin{equation}
  2\sqrt{-g}\tilde E= 
     \partial_\alpha(g^{\alpha\beta}\partial_\beta \gamma^{\mu\nu})
      - \tilde S^{\mu\nu} =0
    \label{eq:fc}
\end{equation}
where $\tilde S^{\mu\nu}$ is comprised of (nonlinear) first derivative
terms in (\ref{eq:etilde}) that do not enter the principle part.

Numerical evolution is implemented on a spatial
grid  $(x_{I}, y_{J}, z_{K})=(Ih, \, Jh, \, Kh)$, $0\le (I,J,K) \le N$,
with uniform spacing $h$, 
on which a field $f(t,x^i)$ is
represented by its grid values $f_{[I,J,K]}(t) = f(t, x_{I}, y_{J}, z_{K})$. 
The time integration is carried out by the method of lines using a 4th
order Runge-Kutta method. 
We introduce the standard finite difference operators $D_{0i}$
and $D_{\pm i}$ according to the examples 
\begin{eqnarray}
    D_{0x} f_{I,J}&=&\frac{1}{2h}(f_{I+1,J}-f_{I-1,J}) \nonumber \\
    D_{+x} f_{I,J}&=&\frac{1}{h}(f_{I+1,J}-f_{I,J}) \nonumber \\
   D_{-y} f_{I,J}&=&\frac{1}{h}(f_{I,J}-f_{I,J-1}) \nonumber \, ;
\end{eqnarray}
the translation operators $T_{\pm i}$ according to the
example 
\begin{equation}
        T_{\pm x}f_{I,J}=f_{I\pm 1,J}  
\end{equation}
and the averaging operators $A_{\pm i}$ and  $A_{0i}$ ,
according to the examples
\begin{eqnarray}
     A_{\pm x}f_{I,J} = \frac{1}{2}(T_{\pm x}+1)f_{I,J} \nonumber 
\end{eqnarray}
\begin{eqnarray}
     A_{0x}f_{I,J} = \frac{1}{2}(T_{+x}+T_{-x})f_{I,J}\nonumber .
\end{eqnarray}
Standard centered differences $D_{0i}$ are used to approximate the first
derivative terms in (\ref{eq:fc}) comprising $\tilde S^{\mu\nu}$, except at
boundary points where  one-sided derivatives are used when necessary.

We will describe the finite difference techniques for the principle part of
(\ref{eq:fc}) in terms of the scalar wave equation 
\begin{equation}
    \partial_\alpha(g^{\alpha\beta}\partial_\beta \Phi ) =0.
    \label{eq:scalarwave}
\end{equation}
The principle part of the linearization of (\ref{eq:fc})  gives rise to
(\ref{eq:scalarwave}) for each component of $\gamma^{\mu\nu}$.  The
non-vanishing shift introduces mixed space-time derivatives
$\partial_t\partial_i$ which complicates the use of standard explicit
algorithms for the wave equation. This problem has been addressed
in~\cite{alcsch,calab,excis,calabgund}. Here we base our work on
evolution-boundary algorithms shown to be stable for a model 1-dimensional wave
equation with shift~\cite{excis}. The algorithms have the summation by parts
(SBP) property that gives rise to an energy estimate for (\ref{eq:scalarwave})
provided the metric $g^{ij}$ is positive definite, as is the case when the
shift is subluminal (i.e. when the evolution direction
$t^\alpha\partial_\alpha=\partial_t$ is timelike). Alternative explicit finite
difference algorithms are available which are stable for a superluminal
shift~\cite{calab,excis}. Although such superluminal algorithms have importance
in treating the region inside a black hole, they have awkward properties at a
timelike boundary and are not employed in the test cases treated here.

The algorithm we use is designed to obey the semi-discrete versions of two
conservation laws of the continuum system (\ref{eq:scalarwave}).
These govern the monopole quantity
\begin{equation}
     Q=-\int_V g^{t\alpha}\Phi_{,\alpha} dV
\label{eq:contQ}
\end{equation}
and the energy 
\begin{equation}
    E=\frac{1}{2}\int_V (-g^{tt}\Phi_{,t}^2 +g^{ij}\Phi_{,i} \Phi_{,j})dV,
\label{eq:contE}
\end{equation}
where $dV=dxdydz$. By assumption, the $t=const$ Cauchy hypersurfaces are
spacelike so that $-g^{tt}>0$. We also assume in the following that $g^{ij}$ is
positive definite (subluminal shift) so that $E$ provides a norm.
Note that in the gravitational case there are 10 quantities $Q^{\alpha\beta}$
corresponding to (\ref{eq:contQ}), which have monopole, dipole or quadrupole
transformation properties depending on the choice of indices.

For periodic boundary conditions (or in the absence of a boundary),
(\ref{eq:scalarwave}) implies strict monopole conservation $Q_{,t} =0$; and,
when the coefficients of the wave operator are frozen in time, i.e. when
$\partial_t g^{\alpha\beta}=0$, (\ref{eq:scalarwave}) implies strict
energy conservation $E_{,t}=0$. In the time-dependent, boundary-free case,
\begin{equation}
       E_{,t}=\frac{1}{2} \int_V ( g^{\alpha\beta}_{,t}\Phi_{,\alpha}
                             \Phi_{,\beta} ) dV.
\end{equation}
which readily provides an estimate of the form
\begin{equation}
       E_{,t} < k E
\end{equation}
for some $k$ independent of the initial data for $\Phi$. Thus
the norm  is bounded relative to its initial value at $t=0$ by
\begin{equation}
       E < E_0 e^{kt}.
\label{eq:est}
\end{equation}

More generally, these conservation laws
include flux contributions from the boundary,
\begin{equation}
     Q_{,t}=\oint_{\partial V} N_i g^{i\alpha}\Phi_{,\alpha} dS,
\label{eq:fcontQ}
\end{equation}
and 
\begin{equation}
     E_{,t}=-\oint_{\partial V}{\cal F} dS,
\label{eq:fcontE}
\end{equation}
where
\begin{equation}
   {\cal F} =-\Phi_{,t} g^{i\alpha}N_i \Phi_{,\alpha}
\end{equation}
and where the surface area element $dS$ and the unit outward normal $N_i=$ are
normalized with the Euclidean metric $\delta_{ij}$. 
The monopole flux vanishes for the homogeneous Neumann boundary condition
\begin{equation}
   g^{i\alpha}N_i \Phi_{,\alpha}=0.
   \label{eq:hneum}
\end{equation} 
The energy flux is dissipative for boundary conditions such that  ${\cal F}\ge
0$, e.g. a homogeneous Neumann boundary condition or a homogeneous
Dirichlet boundary condition $\Phi_{,t}=0$ or superpositions $g^{i\alpha}N_i
\Phi_{,\alpha} +k\Phi_{,t}=0$ with $k>0$.

It is sufficient to confine the description of the finite difference
techniques to the $(x,y)$ plane and, for brevity, we present the evolution
and boundary algorithms in terms of the 2-dimensional version of
(\ref{eq:scalarwave}) .

\subsection{The evolution algorithm}
\label{sec:iev}

We now discuss the evolution algorithm in the 2-dimensional case with periodic
boundary conditions $f_{0,J}=f_{N,J}$ and $f_{I,0}=f_{I,N}$. We define the
semi-discrete versions of (\ref{eq:contQ}) and (\ref{eq:contE}) as
\begin{equation}
   Q=h^2\sum_{(I,J)=1}^N (-g^{tt}\Phi_{,t} - g^{ti} D_{0i}\Phi).
\label{eq:perq}
\end{equation}
and
\begin{equation}
   E=h^2\sum_{(I,J)=1}^N{\cal E},
\label{eq:peren}
\end{equation}
where
\begin{eqnarray}
    {\cal E} = &-&\frac{1}{2}g^{tt}\Phi_{,t}^2 
            +\frac{1}{4}(A_{+x}g^{xx})(D_{+x}\Phi)^2
             +\frac{1}{4}(A_{-x}g^{xx})(D_{-x}\Phi)^2 \nonumber \\
             &+&\frac{1}{4}(A_{+y}g^{yy})(D_{+y}\Phi)^2
             +\frac{1}{4}(A_{-y}g^{yy})(D_{-y}\Phi)^2 \nonumber \\
             &+&g^{xy}(D_{0x}\Phi)D_{0y}\Phi  .
\label{eq:caleB}    
\end{eqnarray}
The energy $E$ provides a norm on the discretized system, i.e. $E=0$ implies
$\Phi_{,t}=D_{\pm i} \Phi=0$ (provided $g^{ij}$ is positive definite and the
grid spacing is sufficiently small to justify the continuum inequalities
resulting from positive-definiteness).

The simplest second order approximation to (\ref{eq:scalarwave}) which reduces
in the 1-dimensional case to the SBP algorithm presented in~\cite{excis} is 
\begin{equation}
   W:= - \partial_t(g^{tt} \partial_t\Phi) 
        -\partial_t(g^{ti}D_{0i}\Phi)
         - D_{0i}(g^{it} \partial_t\Phi)
                 -{\cal D}_g^2 \Phi =0
\label{eq:2wB}
\end{equation}
where
\begin{eqnarray}
      {\cal D}_g^2 \Phi  &=& 
            \frac{1}{2}D_{+x}\bigg((A_{-x}g^{xx}) D_{-x}\Phi \bigg)
          +\frac{1}{2}D_{-x}\bigg((A_{+x}g^{xx}) D_{+x}\Phi \bigg)
                                     \nonumber \\
         &+& \frac{1}{2}D_{+y}\bigg((A_{-y}g^{yy}) D_{-y}\Phi \bigg)
          +\frac{1}{2}D_{-y}\bigg((A_{+y}g^{yy}) D_{+y}\Phi \bigg)
                                     \nonumber \\
          &+& D_{0x}\bigg(g^{xy} D_{0y}\Phi \bigg)
           + D_{0y}\bigg(g^{xy} D_{0x}\Phi \bigg)  .   
\label{eq:cald}
\end{eqnarray}
The semi-discrete conservation law $Q_{,t}=0$ for
the case of periodic boundary conditions follows immediately
from the flux conservative form of $W$. In order to establish the SBP
property we consider the frozen coefficient case $\partial_t
g^{\alpha\beta}=0$. Then a straightforward calculation gives
\begin{eqnarray}
    {\cal E}_{,t}-\Phi_{,t} W &=& \frac{1}{2} D_{+i}\bigg( 
          g^{ti}\Phi_{,t} T_{-i}\Phi_{,t}+ \Phi_{,t} T_{-i}(g^{ti}\Phi_{,t})
                                   \bigg)\nonumber \\
  &+&\frac{1}{2} D_{+x}\bigg( (A_{-x}g^{xx})(D_{-x}\Phi)T_{-x}\Phi_t \bigg)
    +\frac{1}{2} D_{-x}\bigg( (A_{+x}g^{xx})(D_{+x}\Phi)T_{+x}\Phi_t \bigg)
                   \nonumber \\
  &+&\frac{1}{2} D_{+y}\bigg( (A_{-y}g^{yy})(D_{-y}\Phi)T_{-y}\Phi_{,t} \bigg)
   +\frac{1}{2} D_{-y}\bigg( (A_{+y}g^{yy})(D_{+y}\Phi)T_{+y}\Phi_{,t} \bigg)
                   \nonumber \\   
   &+&\frac{1}{2} D_{+x}\bigg( \Phi_t T_{-x}(g^{xy}D_{0y}\Phi)
                     +g^{xy}(D_{0y}\Phi) T_{-x}\Phi_{,t} \bigg)
                   \nonumber \\   
   &+&\frac{1}{2} D_{+y}\bigg( \Phi_tT_{-y}(g^{xy}D_{0x}\Phi)
                     +g^{xy}(D_{0x}\Phi)  T_{-y}\Phi_{,t} \bigg) .
\label{eq:caledotB}    
\end{eqnarray}

Because each term in (\ref{eq:caledotB}) is a total $D_{\pm i}$, the
semi-discrete conservation law $E_{,t}=0$ follows (for periodic boundary
conditions) when the evolution algorithm $W=0$ is satisfied. When the
coefficients of the wave operator are time dependent, an energy estimate can be
established analogous to (\ref{eq:est}) for the continuum case.

We also consider a modification of the algorithm (\ref{eq:est}) by
introducing extra averaging operators according to
\begin{equation}
  \hat W := - \partial_t(g^{tt} \partial_t\Phi) 
        -\partial_t\bigg( g^{ti}D_{0i}\Phi  \bigg) 
         -  D_{-i}\bigg( (A_{+i}g^{it})(A_{+i}\partial_t\Phi) \bigg)     
	    - D_g^2 \Phi =0 ,
\label{eq:2wa}
\end{equation}
with
\begin{eqnarray}
      D_g^2 \Phi  &=& 
            \frac{1}{2}D_{+x}\bigg((A_{-x}g^{xx}) D_{-x}\Phi \bigg)
          +\frac{1}{2}D_{-x}\bigg((A_{+x}g^{xx}) D_{+x}\Phi \bigg)
                                     \nonumber \\
         &+& \frac{1}{2}D_{+y}\bigg((A_{-y}g^{yy}) D_{-y}\Phi \bigg)
          +\frac{1}{2}D_{-y}\bigg((A_{+y}g^{yy}) D_{+y}\Phi \bigg)
                                     \nonumber \\
          &+& D_{-x}\bigg( (A_{+x}g^{xy}) (A_{+x}D_{0y}\Phi )\bigg)
           + D_{-y}\bigg((A_{+y}g^{xy})(A_{+y} D_{0x}\Phi \bigg) .    
\label{eq:calda}
\end{eqnarray}
It is easy to verify that $\hat W=W+O(h^2)$ and that both are constructed from
the same stencil of points. Although $\hat W$ does not obey the exact SBP
property with respect to the energy (\ref{eq:caleB}), the experiments
in~\cite{babev} show that it leads to significantly better performance than $W$
for gauge wave tests with periodic boundary conditions.

For the time discretization, we apply the method of lines to the 
large system of ordinary differential equations
\begin{equation}
     {\bf \Phi}_{,tt}=\frac{1}{h}{\bf A \Phi}_{,t}+\frac{1}{h^2}{\bf B \Phi}.
\end{equation}
obtained from the spatial discretization.
Introducing 
\begin{equation}
   {\bf \Phi}_{,t}=\frac{1}{h}{\bf T}, 
\end{equation}
we obtain the first order system
\begin{equation}
\left(
\begin{array}{c}
{\bf T} \\
{\bf \Phi}    
\end{array}
\right)_{,t} =  \frac{1}{h}
\left( 
\begin{array}{cc}
{\bf B} & {\bf A} \\
{\bf I} & {\bf 0}   
\end{array}
\right)  
\left( 
\begin{array}{c}
{\bf T} \\
{\bf V}    
\end{array}
\right) . 
\label{eq:fo}
\end{equation}
We solve this system numerically using a 4th order
Runge-Kutta time integrator. 

\subsection{The boundary algorithm}

Again it suffices to describe the algorithm in the 2-dimensional case.
In the absence of periodic symmetry, we modify the definitions of the
monopole quantity (\ref{eq:perq}) and the energy (\ref{eq:peren}) to include
contributions from the cells at the
$ edges=\{(0,J),(N,J),(I,0),(I,N)\}$ and
$ corners=\{(0,0),(N,0),(0,N),(N,N)\}$
of the rectangular grid:
\begin{equation}
   Q=h^2\sum_{(I,J)=1}^{N-1} (-g^{tt}\Phi_t - g^{ti} D_{0i}\Phi)
       +\sum_{edges} Q_{edges}
             +\sum_{corners}Q_{corner},
\label{eq:q}
\end{equation}
and
\begin{equation}
   E=h^2\sum_{(I,J)=1}^{N-1}{\cal E} +\sum_{edges}E_{edges}
             +\sum_{corners}E_{corners},
\label{eq:en}
\end{equation}
with ${\cal E}$ given again by (\ref{eq:caleB}). 
We carry out the analysis of the semi-discrete conservation
laws using the $W$-algorithm (\ref{eq:2wB}).

We can isolate the contributions from the edges by retaining
periodicity in the $y$-direction, so that the boundary consists of two
circular edges (with no corners) at $I=0$ and $I=N$. 
We assign the contribution 
\begin{equation}
   Q_{I=N}=\frac{h^2}{2}\sum_{J=1}^{N}\bigg( -g^{tt}\Phi_t 
               -g^{tx}D_{-x}\Phi -g^{ty}D_{0y}\Phi \bigg)
\label{eq:qedgeB}    
\end{equation} 
to (\ref{eq:q}) from the $I=N$ boundary, with the analogous contribution from
$I=0$. Assuming that $W=0$ at all interior points, a straightforward
calculation yields
\begin{equation}
     Q_{,t}= h\sum_{J=1}^{N} (\frac{h}{2}W_{0,J}
                  -{\cal N}_{0,J}+\frac{h}{2}W_{N,J}+{\cal N}_{N,J})
\end{equation}
where
\begin{eqnarray}
   {\cal N}&=&  \frac{1}{2}\bigg((A_{+x}g^{xx})D_{+x}\Phi
            +(A_{-x}g^{xx})D_{-x}\Phi +(A_{+x}g^{tx})T_{+x}\Phi_t
            +(A_{-x}g^{tx})T_{-x}\Phi_t \nonumber \\
	   &+&g^{xy}D_{0y}\Phi+A_{0x}(g^{xy}D_{0y}\Phi) \bigg)
	     + \frac{h^2}{4}\bigg( D_{0y}(g^{xy}D_{+x}D_{-x}\Phi)
	    +(\partial_t g^{tx})D_{+x} D_{-x} \Phi \bigg ).
\label{eq:neum}
\end{eqnarray}

Here, ${\cal N}_N=0$ is a second order accurate approximation
to the Neumann condition
(\ref{eq:hneum}), evaluated at the $I=N$ boundary. Because  ${\cal N}_N$
involves {\em ghost} points outside the computational grid, the Neumann
boundary condition must be implemented in the form
\begin{eqnarray}
   \frac{h}{2}W_{N,J}+{\cal N}_{N,J}&=&0 \nonumber \\
    \frac{h}{2}W_{0,J}-{\cal N}_{0,J}&=&0 ,
\label{eq:2ineum}
\end{eqnarray}
which, by the above construction, involve only interior and boundary points.
(This is why the term $D_{0y}(g^{xy}D_{+x}D_{-x}\Phi)$, which integrates to 0
on the boundary, is included in (\ref{eq:neum}).)  After reducing
($\ref{eq:2ineum}$) to first order in time form as in (\ref{eq:fo}), it
provides the Neumann update algorithm for the boundary points when the interior
is evolved using the $W$ algorithm (\ref{eq:cald}).  Correspondingly, when
applying the ${\hat W}$ algorithm (\ref{eq:2wa}) to the IBVP, we update the
boundary using
\begin{eqnarray}
   \frac{h}{2}{\hat W}_{N,J}+{\hat {\cal N}}_{N,J}&=&0 \nonumber \\
    \frac{h}{2}{\hat W}_{0,J}-{\hat {\cal N}}_{0,J}&=&0  
\label{eq:hat2ineum}
\end{eqnarray}
where now 
\begin{eqnarray}
   {\hat {\cal N}}&=&  \frac{1}{2}\bigg((A_{+x}g^{xx})D_{+x}\Phi
            +(A_{-x}g^{xx})D_{-x}\Phi 
           +\frac{1}{2}(g^{tx}+A_{+x}g^{tx})T_{+x}\Phi_t
            +\frac{1}{2}(g^{tx}+A_{-x}g^{tx})T_{-x}\Phi_t \nonumber \\
	   &+&(A_{+x}g^{xy})A_{+x}D_{0y}\Phi
            +(A_{-x}g^{xy})A_{-x}D_{0y}\Phi  \bigg ) \nonumber \\
	  &+& \frac{h^2}{4}\bigg( D_{-y}[(A_{+y}g^{xy})A_{+y}D_{+x}D_{-x}\Phi]
	    +(\partial_t g^{tx})D_{+x} D_{-x} \Phi
            +\frac{1}{2}(D_{+x} D_{-x}  g^{tx})\Phi_t  \bigg ).
\label{eq:hatneum}
\end{eqnarray}

In the case of
inhomogeneous Neumann boundary data $q$, (\ref{eq:2ineum}) takes the form
\begin{eqnarray}
   \frac{h}{2}W_{N,J}+{\cal N}_{N,J}&=&g^{xx}q_N \nonumber \\
    \frac{h}{2}W_{0,J}-{\cal N}_{0,J}&=&-g^{xx}q_0 .
\label{eq:q2ineum}
\end{eqnarray}

We assign the contribution to the energy from the $I=N$ boundary to be
\begin{eqnarray}
   E_{I=N}&=&\frac{h^2}{2}\sum_{J=1}^{N}\bigg(
       -\frac{1}{2}g^{tt}\Phi_t^2 
             +\frac{1}{2}(A_{-x}g^{xx})(D_{-x}\Phi)^2 \nonumber \\
             &+&\frac{1}{4}(A_{+y}g^{yy})(D_{+y}\Phi)^2
             +\frac{1}{4}(A_{-y}g^{yy})(D_{-y}\Phi)^2 \nonumber \\
             &+&g^{xy}(D_{-x}\Phi)D_{0y}\Phi \bigg),
\label{eq:eedgeB}    
\end{eqnarray} 
with the analogous contribution from $I=0$. Assuming that $W=0$ at all
interior points and that the coefficients of the wave operator are
time independent,  this leads to the discrete energy-flux conservation law
\begin{equation}
     E_{,t}= -h\sum_{J=1}^{N} ({\cal F}_{0,J}+{\cal F}_{N,J})
\end{equation}
with
\begin{equation}
            {\cal F}_{N,J} = -\bigg(\Phi_t (\frac{h}{2}W+{\cal N}) \bigg)_{N,J}
\end{equation}
and
\begin{equation}
            {\cal F}_{0,J} = \bigg(\Phi_t (\frac{h}{2}W-{\cal N}) \bigg)_{0,J}.
\end{equation}
Any discrete boundary conditions such that ${\cal F}_{0,J}$ and
${\cal F}_{N,J}$ are non-negative retain the SBP property of the $W$-algorithm.
In particular, this holds for the Dirichlet or Neumann boundary conditions,
for which ${\cal F}=0$ in the homogeneous case. We also consider the
Sommerfeld-type algorithm (with Sommerfeld data $q$).

\begin{eqnarray}
   \bigg( \frac{h}{2}W+{\cal N} 
       +\sqrt{\frac {g^{xx}}{-g_{tt}}}\Phi_t\bigg )_{N,J}&=&q_N \nonumber \\
    \bigg( \frac{h}{2}W-{\cal N}
           +\sqrt{\frac {g^{xx}}{-g_{tt}}} \Phi_t\bigg )_{0,J}&=&q_0,
\label{eq:somm}
\end{eqnarray}
for which ${\cal F}$ is strictly positive in the case of homogeneous data.

The algorithm for the corners follow from the same SBP approach. Since we
restrict our applications in Sec.~\ref{sec:tests} to  tests with smooth
boundaries, we will not present the details.

\subsection{Dissipation}

It is simplest to add dissipation by using the Euclidean Laplacian with
the centered approximation
\begin{equation}
       {\cal D}_e^2 = D_{+x} D_{-x}+D_{+y}D_{-y}+D_{+z}D_{-z} .
\end{equation}
Then, for the 2-dimensional case
with periodic boundary conditions, the modification 
\begin{equation}
       W \rightarrow  W + \epsilon h^3{\cal D}_e^2 {\cal D}_e^2 \Phi_{,t} 
\end{equation}
of the $W$ evolution algorithm (\ref{eq:2wB}) dissipates the energy
(\ref{eq:peren}) according to
\begin{equation}
   E_{,t} \rightarrow  E_{,t} +\epsilon
               h^5\sum_{(I,J)=1}^N ({\cal D}_e^2 \Phi_{,t} )^2 . 
      \label{eq:pediss}
\end{equation}

The dissipation (\ref{eq:pediss}) leaves undisturbed the semi-discrete energy
estimate governing the stability of the $W$ evolution algorithm. We present a
similar SBP result for the boundary algorithm. It suffices to illustrate the
approach in the 1-dimensional case. It is the semi-discrete analogue of adding
a dissipative term to the wave equation, in the half-plane $0\le x$, in the
symbolic form 
\begin{equation}
 \partial_t (-g^{tt} \Phi_{,t}) \rightarrow 
           \partial_t (-g^{tt} \Phi_{,t})
       +\epsilon \{\partial_x^4 \Phi_{,t}
         +\delta(x)\partial_x^3 \Phi_{,t} 
	+\partial_x (\delta (x)\partial_x^2 \Phi_{,t}) \}
	\label{eq:cdiss}
\end{equation}
so that the Dirac delta function terms $\delta(x)$ cancel the boundary
contributions at $x=0$ to give
\begin{equation}
   \int_0^\infty dx\Phi_{,t}\partial_t (-g^{tt} \Phi_{,t})  \rightarrow  
     \int_0^\infty dx \bigg (  \Phi_{,t}\partial_t (-g^{tt} \Phi_{,t}) 
      + \epsilon (\partial_x^2 \Phi_{,t})^2 \bigg) .
\end{equation}
As a result, the energy dissipates according to
\begin{equation}
 \partial_t E \rightarrow \partial_t E  + \epsilon \int_0^\infty dx  
     (\partial_x^2 \Phi_{,t})^2  .
\end{equation}

In order to model this dissipation at the finite difference level, we use
summation by parts techniques. A comprehensive discussion has been given
in~\cite{multiblock}. For the second order accurate approximation to the wave
equation considered here, the following treatment suffices.

On the interval $0\le x_I =Ih \le 1= Nh$ we use the identity
\begin{equation}
   h\sum_2^{N-2}\{fD_-g+(D_+ f) g \}_I=f_{N-1}g_{N-2}-f_2 g_1 
\end{equation}
to obtain
\begin{eqnarray}
  h\sum_2^{N-2} \{ f(D_+D_-)^2 f \}_I  &=&
        - h\sum_2^{N-2} \{(D_- f)D_-^2D_+ f  \}_I
      +(D_-^2D_+f)_{N-1}f_{N-2}  -(D_-^2 D_+f)_2 f_1    \nonumber  \\
	 &=& h\sum_2^{N-2} \{(D_+D_- f)^2  \}_I
	 -(D_-f)_{N-1}(D_+D_-f)_{N-2} +(D_-f)_2 (D_+D_- f)_1  \nonumber  \\
      &&  +(D_-^2D_+f)_{N-1}f_{N-2} -(D_-^2 D_+f)_2 f_1   
            \nonumber  \\ 
      &=& h\sum_2^{N-2} \{(D_+D_- f)^2  \}_I \nonumber  \\
      && +\frac{1}{h}\bigg (f_{N-2}D_+D_- f_{N-1}-f_{N-1}D_+D_- f_{N-2}
	     +f_{2}D_+D_- f_{1} -f_{1}D_+D_- f_{2} \bigg)    .
\end{eqnarray}

Thus we are led to the finite difference analogue of (\ref{eq:cdiss}) 
\begin{eqnarray}
 \{\partial_t (-g^{tt} \partial_t \Phi)\}_I && 
            \rightarrow    { \partial_t (-g^{tt} \partial_t \Phi)}_I 
      + \epsilon h^3(D_+D_-)^2 \partial_t \Phi _I
                  \, , \quad 3\le I\le N-3 \nonumber \\		  
	 &&	 \rightarrow    { \partial_t (-g^{tt} \partial_t \Phi)}_I 
    + \epsilon \{ h^3(D_+D_-)^2 \partial_t \Phi_{N-2} 
              -hD_+^2 \partial_t\Phi_{N-2} \}
                  \, , \quad I= N-2 \nonumber \\
     && \rightarrow { \partial_t (-g^{tt} \partial_t \Phi)}_I 
      + \epsilon h D_-^2 \partial_t\Phi_{N-1} \}
       \, , \quad I= N-1 \nonumber \\
    &&  \rightarrow \nonumber  { \partial_t (-g^{tt} \partial_t \Phi)}_I 
       \, , \quad I= N \, ,\quad  I=0 \nonumber \\
    &&  \rightarrow \nonumber  { \partial_t (-g^{tt} \partial_t \Phi)}_I
     + \epsilon \{ h^3(D_+D_-)^2 \partial_t \Phi_2 
              -hD_-^2 \partial_t\Phi_{2} \}
                  \, , \quad I= 2 \nonumber \\   
        && \rightarrow { \partial_t (-g^{tt} \partial_t \Phi)}_I 
      + \epsilon h D_+^2 \partial_t\Phi_{1} 
       \, , \quad I= 1 .\nonumber \\
        \label{eq:2diss}
\end{eqnarray} 
This dissipation implies (in the 1D case where $E=h\sum {\cal E}$)
\begin{equation}
    E_{,t} \rightarrow E_{,t} 
          + \epsilon h^4\sum_2^{N-2} (D_+D_- \partial_t \Phi_I)^2 .
\end{equation}

Alternatively, we can extend the algorithm by one more point:
\begin{eqnarray}
 \{\partial_t (-g^{tt} \partial_t \Phi)\}_I && 
            \rightarrow    { \partial_t (-g^{tt} \partial_t \Phi)}_I 
      + \epsilon h^3(D_+D_-)^2 \partial_t \Phi _I
                  \, , \quad 2\le I\le N-2 \nonumber \\		  
	 &&	 \rightarrow    { \partial_t (-g^{tt} \partial_t \Phi)}_I 
    + \epsilon \{ h^3(D_+D_-)^2 \partial_t \Phi_{N-1} 
              -hD_+^2 \partial_t\Phi_{N-1} \}
                  \, , \quad I= N-1 \nonumber \\
     && \rightarrow { \partial_t (-g^{tt} \partial_t \Phi)}_I 
      + \epsilon h D_-^2 \partial_t\Phi_N \}
       \, , \quad I= N \nonumber \\
    &&  \rightarrow \nonumber  { \partial_t (-g^{tt} \partial_t \Phi)}_I
     + \epsilon \{ h^3(D_+D_-)^2 \partial_t \Phi_1 
              -hD_-^2 \partial_t\Phi_{1} \}
                  \, , \quad I= 1 \nonumber \\   
        && \rightarrow { \partial_t (-g^{tt} \partial_t \Phi)}_I 
      + \epsilon h D_+^2 \partial_t\Phi_0 
       \, , \quad I= 0 ,
        \label{eq:a2diss}
\end{eqnarray} 
with the dissipation now implying (in the 1D case)
\begin{equation}
    E_{,t} \rightarrow E_{,t} 
          + \epsilon h^4\sum_1^{N-1} (D_+D_- \partial_t \Phi_I)^2.
\end{equation}

Additional dissipation can be added at the boundary by using
the identities
\begin{equation}
       h(D_-f_N)^2=f_N D_-f_N -f_{N-1} D_+f_{N-1}
\end{equation}
and
\begin{equation}
       h(D_+f_0)^2=-f_0 D_+f_0 +f_{1} D_-f_{1}
\end{equation}
to augment the dissipative terms in (\ref{eq:2diss}) by  
\begin{eqnarray}
 \{\partial_t (-g^{tt} \partial_t \Phi)\}_I && 
       	 \rightarrow    { \partial_t (-g^{tt} \partial_t \Phi)}_I 
    - \epsilon_B hD_+ \partial_t \Phi_{N-1} 
                  \, , \quad I= N-1 \nonumber \\
     && \rightarrow { \partial_t (-g^{tt} \partial_t \Phi)}_I 
      + \epsilon_B h D_- \partial_t\Phi_N
       \, , \quad I= N \nonumber \\
    &&  \rightarrow \nonumber  { \partial_t (-g^{tt} \partial_t \Phi)}_I
     + \epsilon_B  hD_-\partial_t \Phi_1 
                  \, , \quad I= 1 \nonumber \\   
        && \rightarrow { \partial_t (-g^{tt} \partial_t \Phi)}_I 
      -\epsilon_B h D_+ \partial_t\Phi_0 
       \, , \quad I= 0 .\nonumber \\
        \label{eq:bdiss}
\end{eqnarray} 
This results in the dissipative effect on the energy (in the 1D case)
\begin{equation}
    E_{,t} \rightarrow E_{,t} 
          + \epsilon_B h^2 (D_+ \partial_t \Phi_0)^2
	  + \epsilon_B h^2 (D_- \partial_t \Phi_N)^2 .
\end{equation}

\subsection{Implementing the constraint-preserving boundary system}
\label{sec:cpbs}

For a boundary located at $x=const$, we have shown that the
evolution-boundary algorithm is constraint-preserving if the boundary data
(\ref{eq:qdata}) are provided and the remaining boundary values
are updated as follows:

\begin{enumerate}

\item The Dirichlet boundary data $q^a=\gamma^{xa}/\gamma^{xx}$
is freely prescribed. 

\item The Neumann boundary data $q^{xx}=q^\mu \partial_\mu \gamma^{xx}$ is
determined by (\ref{eq:gqzz}). 

\item The Neumann boundary data $q^{ab}= q^\mu \partial_\mu \gamma^{ab}$ is
computed in terms of three free functions from the boundary system
(\ref{eq:gakconstr}), recast in the symmetric hyperbolic form (\ref{eq:Yeq3}).

\item $\partial_t \gamma^{\mu\nu}$ is updated on the boundary using the
constrained boundary data; then $\gamma^{\mu\nu}$ is updated.

\end{enumerate}

The first two items involve minimal computation. The update algorithm ensures
that the components ${\gamma}^{xx}$ and ${\gamma}^{ab}$ are known at the
current time level. Then the Dirichlet data $q^a$ determines the remaining
components $\gamma^{ax}$ (by item 1) and the Neumann data  $q^{xx}$ (by item
2).

In describing the implementation of the third and fourth items, we consider for
the present purposes a smooth boundary consisting of two toroidal faces at
$x=0$ and $x=1$ (periodic boundary conditions in the $y$ and $z$ directions).
Again, the grid points are labeled by $0\le (I,J,K)\le N$. The updates of the
boundary values $\gamma^{xx}$ and $\gamma^{ab}$ in terms of their Neumann
boundary data is based upon the generalization of the scalar update scheme
(\ref{eq:q2ineum}) to include the non-principle-part terms of the gravitational
equations. For example, the component of (\ref{eq:q2ineum}) corresponding to
$\gamma^{ab}$ generalizes at the $I=N$ boundary to
\begin{equation}
  \bigg( -h\sqrt{-g}\tilde E^{ab}+{\cal N}^{ab} 
          -g^{xx} q^{ab} \bigg)_{(N,J,K)} =0,
\label{eq:3ineum}
\end{equation}
where $\sqrt{-g}\tilde E^{ab}$ is finite differenced according to the
rules given in Sec.~\ref{sec:iev} and where 
\begin{eqnarray}
   {\cal N}^{ab}&=&  \frac{1}{2}\bigg((A_{+x}g^{xx})D_{+x}\gamma^{ab}
        +(A_{-x}g^{xx})D_{-x}\gamma^{ab} +(A_{+x}g^{tx})T_{+x}\gamma^{ab}_{,t}
            +(A_{-x}g^{tx})T_{-x}\gamma^{ab}_{,t} \nonumber \\
	   &+&g^{xy}D_{0y}\gamma^{ab}+A_{0x}(g^{xy}D_{0y}\gamma^{ab}) 
        +g^{xz}D_{0z}\gamma^{ab}+A_{0x}(g^{xz}D_{0z}\gamma^{ab})  \bigg)
               \nonumber \\
	     &+& \frac{h^2}{4}\bigg( D_{0y}(g^{xy}D_{+x}D_{-x}\gamma^{ab})
              +D_{0z}(g^{xz}D_{+x}D_{-x}\gamma^{ab})    
	    +(\partial_t g^{tx})D_{+x} D_{-x} \gamma^{ab} \bigg ),
\label{eq:3neum}
\end{eqnarray}
as in (\ref{eq:neum}).

The constrained Neumann data $q^{ab}$ is obtained from the first order
symmetric hyperbolic boundary system (\ref{eq:Yeq3}), which is solved by the
method of lines using centered spatial differences and a 4th order
Runge-Kutta time integrator. The required  data ${\cal Q}^{AB}$ for this system
(where $x^A=(y,z)$) is related to the boundary data $q^{ab}$ by
(\ref{eq:qfromy}), or its inverse
\begin{equation}
         {\cal Q}^{ab}=q^{ab}-\eta^{ab}\eta_{cd}q^{cd}.
\end{equation}
This gives
\begin{eqnarray}
   {\cal Q}^{zz}-{\cal Q}^{yy}&=& q^{zz}-q^{yy},  \nonumber \\  
   {\cal Q}^{yz}&=& q^{yz},  \nonumber \\  
   {\cal Q}^{zz}+{\cal Q}^{yy}&=&2q^{tt}- (q^{zz}+q^{yy}).
\label{eq:freed}
\end{eqnarray}
Integration of (\ref{eq:Yeq3}) then determines the constrained data
\begin{eqnarray}
   \phi=\frac{1}{2}{\cal Q}^{tt}&=& \frac{1}{2} (q^{zz}+q^{yy})  \nonumber \\  
   {\cal Q}^y={\cal Q}^{ty} &=& q^{ty}  \nonumber \\  
   {\cal Q}^z={\cal Q}^{tz}&=& q^{tz}.
\label{eq:constrd}
\end{eqnarray}

In the update of the Neumann components ${\gamma}^{ab}$ via (\ref{eq:3ineum}),
the inhomogeneous term $g^{xx}q^{ab}$ introduces numerical error (from the
evolution of $g^{xx}$) which effects the exact nature of the semi-discrete
multipole conservation laws satisfied by the principle part of the system. We
avoid this source of error for the components corresponding to $Q^{AB}$ by
prescribing data for $g^{zz}Q^{AB}$ (rather than $Q^{AB}$); similarly,
we prescribe $g^{zz}q^{zz}$. In this way, the
principle part of the system obeys exact inhomogeneous versions of the
semi-discrete conservation laws for the components corresponding to $Q^{AB}$.

\section{Tests of the evolution-boundary algorithms}
\label{sec:tests}

We compare the performance of various versions
of the evolution-boundary algorithm
using the AppleswithApples gauge wave metric
\begin{equation}
  \label{eq:flatgaugewave4metric}
  ds^2=(1-H)(-dt^2 +dx^2)+dy^2+dz^2,
  \label{eq:gw}
\end{equation}
and a shifted version given by the
Kerr-Schild metric
\begin{equation}
  \label{eq:sgaugewave4metric}
  ds^2=- dt^2 +dx^2+dy^2+dz^2 +H k_\alpha k_\beta dx^\alpha dx^\beta,
  \label{eq:sgw}
\end{equation}
where, in both cases,
\begin{equation}
  \label{eq:flatgaugewaveHfn}
  H = H(x-t)= A \sin \left( \frac{2 \pi (x - t)}{d} \right),
\end{equation}
and
\begin{equation}
      k_\alpha=\partial_\alpha (x-t)=(-1,1,0,0).
\end{equation}
These metrics describes sinusoidal gauge waves of amplitude $A$ propagating
along the $x$-axis. In order to test 2-dimensional features, 
we rotate the coordinates according to 
\begin{equation}
  \label{eq:flatgaugewave1to2d}
   x = \frac{1}{\sqrt{2}}(x^\prime - y^\prime), \qquad
    y = \frac{1}{\sqrt{2}}(x^\prime + y^\prime) \, .
\end{equation}
which produces a gauge wave propagating along the diagonal. 

The results of these gauge wave tests for periodic boundaries, i.e. a 3-torus
$T^3$ without boundary, have been reported and discussed in~\cite{babev}. In
the periodic case, it was found that the 1D tests were very discriminating in
revealing problems. The 2D tests were essential for designing algorithms for
handling the mixed second derivatives in the  Laplacian operator but they
introduced no new instabilities of an analytic origin. As a result, in the 1D
tests, numerical error is channeled more effectively into exciting nonlinear
instabilities of the analytic problem. 

Here we open up the $x$-axis of the 3-torus to form a manifold with  $T^2$
boundaries at $x=\pm.5$.  Most of our tests are run, in both axis-aligned and
diagonal form, with amplitude $A=0.5$. We have found that the smaller
amplitudes, $A=.01$ and $A=.1$, of the original AwA specifications are not as
efficient for revealing problems. Larger amplitudes can trigger gauge
pathologies more quickly, e.g $g^{tt}\ge 0$ (breakdown of the spacelike nature
of the Cauchy hypersurfaces). Otherwise, the tests are carried out with the
original AwA specifications. We choose wavelength $d=1$ in the axis-aligned
simulation and wavelength $d'=\sqrt{2}$ in the diagonal simulation and make the
following choices for the computational grid:
\begin{itemize}
\item Simulation domain:
\begin{center}
\begin{tabular}{rllll}
  1D:& $\quad  x \in [-0.5, +0.5],$ & $\quad y = 0,$ &$ \quad z = 0,$
  & $ \quad d=1$ \\
  diagonal: & $\quad x \in [-0.5, +0.5], $ & 
    $\quad y \in [-0.5, +0.5], $ & $\quad z=0,$ & $\quad d'=\sqrt{2}$
\end{tabular}
\end{center}
\item Grid: $x_n = -0.5 + n dx, \quad n=0,1\ldots 50\rho,
  \quad dx=dy=dz=1/(50\rho), \quad \rho = 1, 2, 4 $
\item Time step: $dt = dx/4 = 0.005 / \rho$ .
\end{itemize}
The grids have $N=50 \rho =(50,100,200)$ zones. (At least 50 zones are required
to lead to reasonable simulations for more than 10 crossing times.)  The 1D
tests are carried out for $t=1000$ crossing times, i.e.~$2\times10^5\rho$
time steps (or until the code crashes) and the 2D tests for 100 crossing times.

For the case without periodic boundaries in the $x$-direction, the boundary
data are provided at $x=\pm .5$. For example, in the 1D simulation, for a
formulation with a Sommerfeld boundary condition on the metric component
$g_{tt}$ the correct boundary data are 
\begin{eqnarray}
      (\partial_t-\partial_x)g_{tt}|_{x=-.5}  &=& 2\partial_t H(-.5-t)  \\
     (\partial_t+\partial_x)g_{tt}|_{x=.5}  &=& 0.
\end{eqnarray}
With this inhomogeneous Sommerfeld data, the wave enters through the boundary
at $x=-.5$, propagates across the grid and exits through the boundary at
$x=.5$. Inhomogeneous Dirichlet and Neumann data are supplied in the analogous
way. Analytic boundary data are prescribed for all 10 metrics components except
for  the constraint preserving boundary algorithm, where only 6 pieces of
analytic data are provided.

The test results reported here are for the $\hat W$ algorithm (\ref{eq:2wa}),
which performed significantly better than the $W$ algorithm in tests in the
boundary-free case~\cite{babev}. Also, harmonic gauge forcing terms were not
found to be effective in the boundary-free gauge wave tests and we have not not
included them in the present tests. (Gauge forcing is important in spacetimes
where harmonic coordinates are pathological, e.g. the standard $t$-coordinate
in Schwarzschild spacetime is harmonic but singular at the horizon.) 

We use the $\ell_\infty$ norm to measure the error
\begin{equation}
       {\cal E}=||\Phi_\rho-\Phi_{ana}||_\infty
\end{equation}
in a grid function $\Phi_\rho$ with known analytic value $\Phi_{ana}$. We
measure the convergence rate at time $t$
\begin{equation}
    r(t) = \log_2 \big (
   \frac{||\Phi_2-\Phi_{ana}||}{||\Phi_4-\Phi_{ana}||} \big ),
\end{equation}
using the $\rho=2$ and $\rho=4$ grids ($N=100$ and $N=200)$.
It is also convenient to graph the rescaled error
\begin{equation}
      {\cal E_\rho} =\frac{\rho^2}{16}||\Phi_\rho-\Phi_{ana}||_\infty,
\end{equation}
which is normalized to the $\rho=4$ grid.

\subsection{Gauge wave boundary tests}

Simulation of the AwA gauge wave without shift (\ref{eq:gw}) is complicated by
the related metric~\cite{bab}
\begin{equation}
  ds_\lambda^2=e^{\lambda t}(1-H)(- dt^2 +dx^2)+dy^2+dz^2 .
\label{eq:lgw}
\end{equation}
For any value of $\lambda$, this is a flat metric which obeys the harmonic
coordinate conditions and represents a harmonic gauge instability of Minkowski
space with periodic boundary conditions. Since (\ref{eq:lgw}) satisfies the
harmonic conditions, constraint adjustments are ineffective in controlling this
instability. Long term harmonic evolutions of the AwA gauge wave with periodic
boundary conditions were achieved by suppressing this unstable mode by the use
of semi-discrete conservation laws for the principle part of the system. 

These semi-discrete conservation laws have been incorporated in
Sec.~\ref{sec:imp} in the harmonic evolution-boundary algorithm with a general
dissipative boundary condition. Consequently, good long term performance should
continue to be expected. This is especially true for Dirichlet or
Sommerfeld boundary conditions, which do not allow the unstable mode
(\ref{eq:lgw}) at the continuum level. However, Neumann boundary conditions
allow this mode so that it poses a potential problem for the constraint
preserving boundary algorithm described in Sec.~\ref{sec:cpbs}, which combines
Dirichlet and Neumann boundary conditions. Nevertheless, our expectations of
good long term performance are borne out in all our tests of the gauge wave
without shift. No explicit dissipation was added to the $\hat W$-algorithm.

\begin{figure}[hbtp]
  \centering
  \includegraphics*[width=7cm]{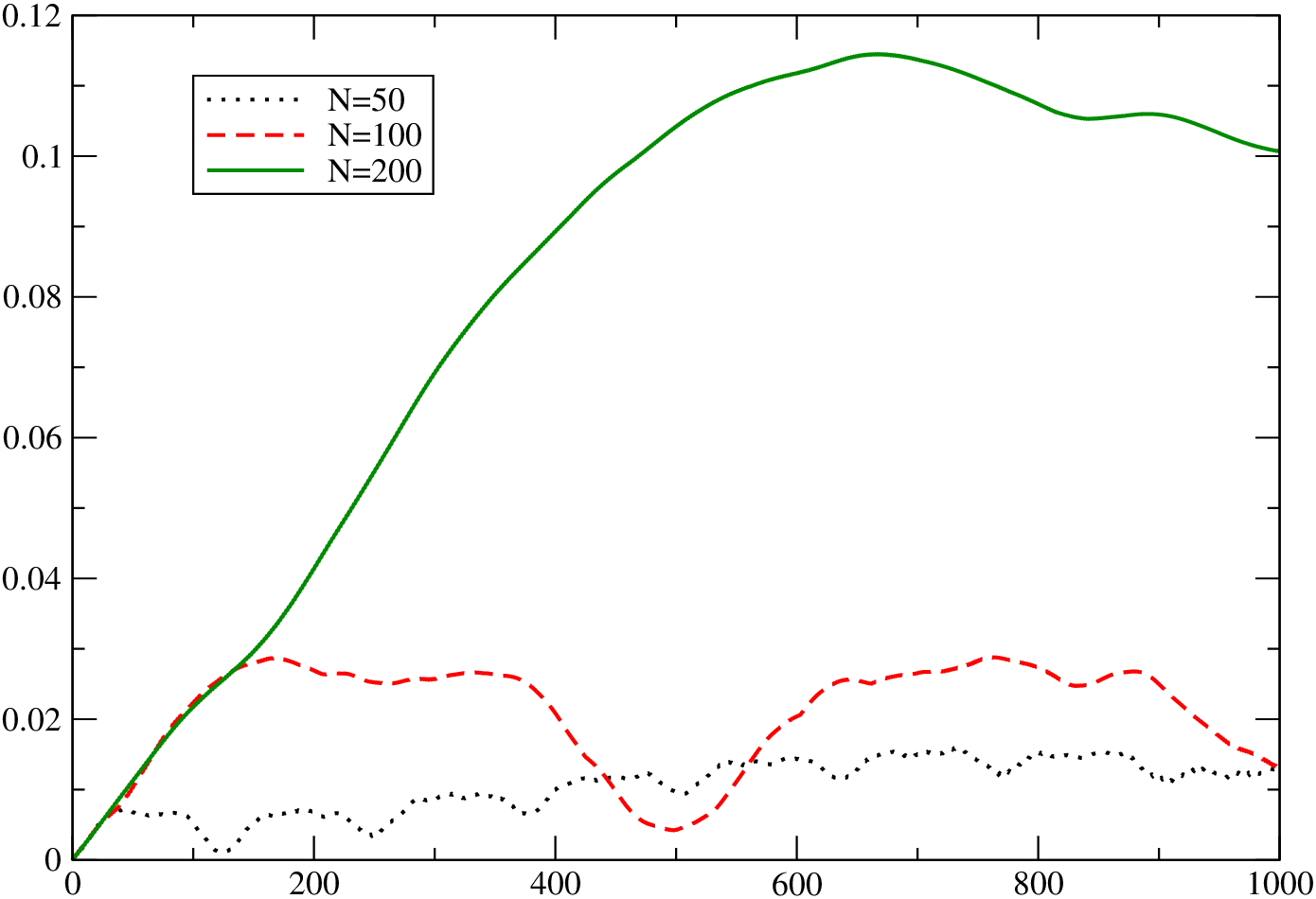}
  \caption{Plot of the rescaled error ${\cal E}_\rho(t)$ in $g_{xx}$
  for the 1D gauge wave with 10 Dirichlet boundary conditions using
  the bare ${\hat W}$ algorithm (no constraint adjustment or dissipation) 
  obtained with $N=50$, 100 and 200 grid zones.
  The horizontal axis measures crossing time $t$. The instabilities are kept
  under control during the entire 1000 crossing times run.}
  
  \label{fig:GW1D10Dconv}
\end{figure}

\begin{figure}[hbtp]
  \centering
  \includegraphics*[width=7cm]{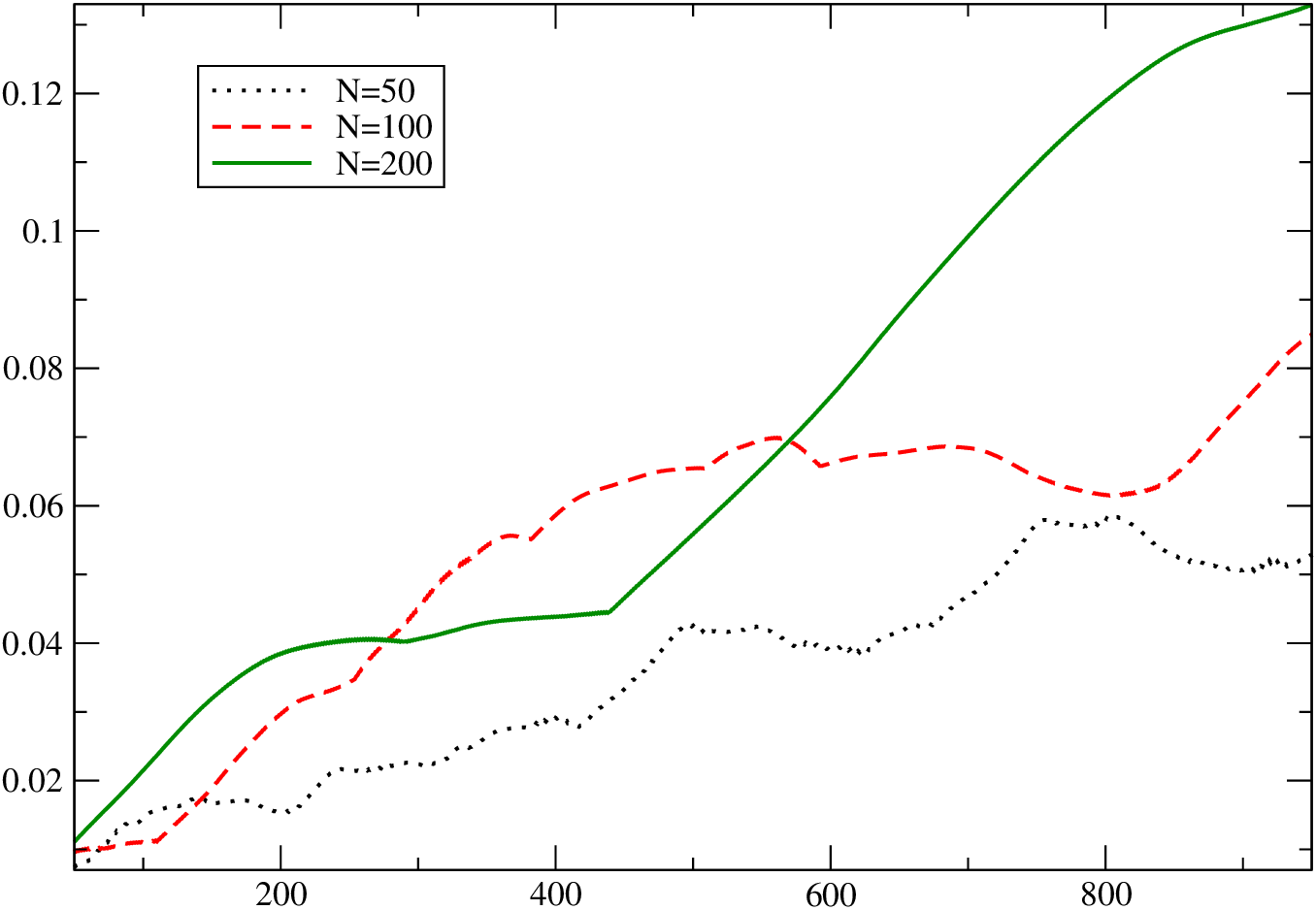}
  \caption{Plot of the rescaled error ${\cal E}_\rho(t)$ in $g_{xx}$
  for the 1D gauge wave as in Fig.~\ref{fig:GW1D10Dconv} but with
  10 Neumann boundary conditions.}
  \label{fig:GW1D10Nconv}
\end{figure}

\begin{figure}[hbtp]
  \centering
  \includegraphics*[width=7cm]{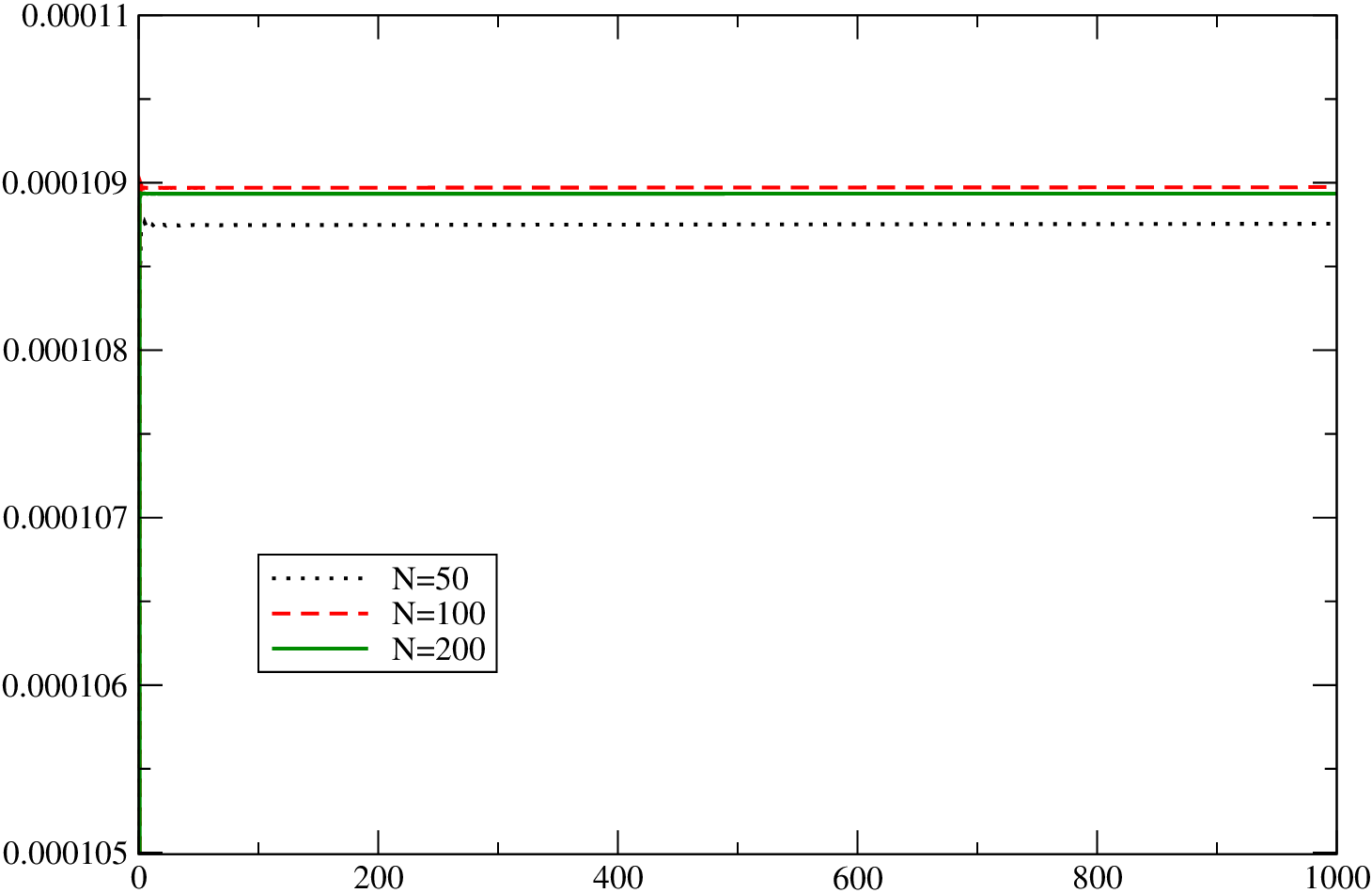}
  \caption{Plot of the rescaled error ${\cal E}_\rho(t)$ in $g_{xx}$
  for the 1D gauge wave as in Fig.~\ref{fig:GW1D10Dconv} but with
  10 Sommerfeld boundary conditions. Excellent performance with
  clean second order convergence is manifest for 1000 crossing times.}
  \label{fig:GW1D10Sconv}
\end{figure}

Figures \ref{fig:GW1D10Dconv}, \ref{fig:GW1D10Nconv} and \ref{fig:GW1D10Sconv}
plot the rescaled error ${\cal E}_\rho(t)$ in $g_{xx}$ for tests with
amplitudes $A=.5$ and either 10 Dirichlet, 10 Neumann or 10 Sommerfeld boundary
conditions. The plots for the Dirichlet and Neumann boundary conditions show
accuracy comparable to the corresponding tests with periodic boundary
conditions reported  in~\cite{babev}. The convergence rates measured at
$t=50$ are $r_{10D}(50) = 1.876$ for the Dirichlet case and $r_{10N}(50) =
1.804$ for the Neumann case. The Dirichlet and Neumann boundary algorithms
supply the correct inhomogeneous boundary data for the gauge wave signal to
leave the grid but the numerical error is reflected at the boundaries and
accumulates in the grid. The Sommerfeld boundary condition allows this
numerical error to leave the grid and gives excellent long term
performance, as evidenced by the clear second order convergence displayed in
Fig.~\ref{fig:GW1D10Sconv} throughout the 1000 crossing time run. The
convergence rate $r_{10S}(50) = 2.000$  was measured at $t=50$.

Test results using 3 Dirichlet boundary conditions (for the components
$\gamma^{xa}$) and seven Neumann boundary conditions (for the components
$\gamma^{xx}$ and $\gamma^{ab}$), with boundary data supplied by the analytical
solution, were practically identical to results for the 10 Neumann boundary
conditions. When these seven pieces of Neumann boundary data was generated by
the constraint preserving boundary system, as detailed in Sec.~\ref{sec:cpbs},
practically identical results were again found. This is because no appreciable
constraint violation is excited in the shift-free gauge wave test. For the same
reason, constraint adjustments have no significant effect.

Two-dimensional tests of the diagonally propagating gauge wave were also in
line with expectations. The most important background information for the
design of physically relevant boundary algorithms using CCM comes from the case
of 10 Sommerfeld boundary conditions and the case of boundary conditions
supplied by the constraint preserving (CP) boundary algorithm.  In 2D tests
of these algorithms, we found the respective convergence rates 
\begin{eqnarray}
       r_{10S}(10) = 2.0005 \, , \quad r_{10S}(100) = 2.0004 \nonumber \\
       r_{CP}(10) = 2.0482 \, , \quad r_{CP}(100) = 1.0338 \nonumber
\end{eqnarray} measured at 10 and 100 crossing times. The Sommerfeld case
maintains clean second order convergence. At 10 crossing times the constraint
preserving system also shows clean second order convergence but at 100 crossing
times the convergence rate drops to first order due to accumulation of error.  
The constraint boundary system (\ref{eq:Yeq3}) involves two derivatives of
numerically evolved quantities in order to provide a complete set of Neumann
data to the code. This introduces high frequency error in the Neumann data. We
have not experimented with ways to dissipate this source of error in the
boundary system.

\subsection{Shifted gauge wave boundary tests}

Simulation of the shifted gauge wave (\ref{eq:sgw}) is complicated by
the related exponentially growing Kerr-Schild metric~\cite{babev}
\begin{equation}
  \label{eq:lsgaugewave4metric}
  ds_\lambda^2=- dt^2 +dx^2+dy^2+dz^2 +
        \bigg(H-1+e^{\lambda \hat t}\bigg )
            k_\alpha k_\beta dx^\alpha dx^\beta ,
\label{eq:instab}
\end{equation}
where
\begin{equation}
\hat t= t - \frac {Ad}{4\pi}\cos \left( \frac{2 \pi (x - t)}{d} \right) .
\end{equation}
Although this metric does not solve Einstein's
equations, for any value of $\lambda$ it satisfies the standard harmonic form
(\ref{eq:e}) of the reduced Einstein equations, i.e. the equations
upon which the numerical evolution is based. Numerical error in the shifted
gauge wave test should be expected to excite this instability. As a result, in
shifted gauge wave tests with periodic boundary conditions~\cite{babev}, the
constraint adjustment (\ref{eq:a1adj}) or (\ref{eq:a2adj}) was necessary to
modify the form of the reduced system in order to eliminate the solution
(\ref{eq:instab}).  

\begin{figure}[hbtp]
  \centering
   \includegraphics*[width=7cm]{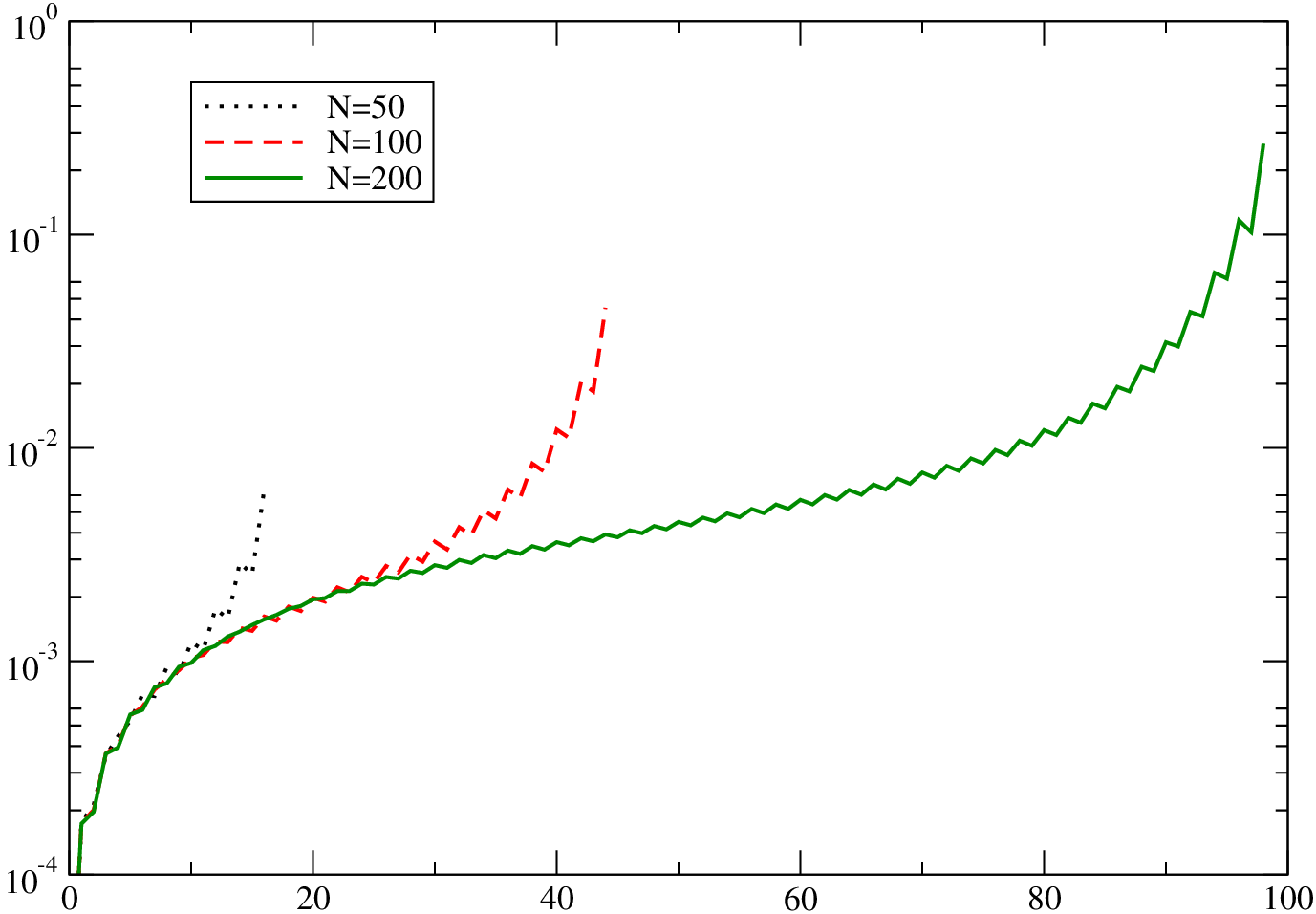}
  \caption{Log plot of the rescaled error ${\cal E}_\rho(t)$ in $g_{xx}$
  for the 1D gauge wave with shift for the bare $\hat W$ algorithm
  with 10 Dirichlet boundary conditions. Although clean second order convergence
  is measured at early times, the simulation goes unstable in less than 100
  crossing times.} 
  \label{fig:GW1DSH10Dconv}
\end{figure}

\begin{figure}[hbtp]
  \centering
  \includegraphics*[width=7cm]{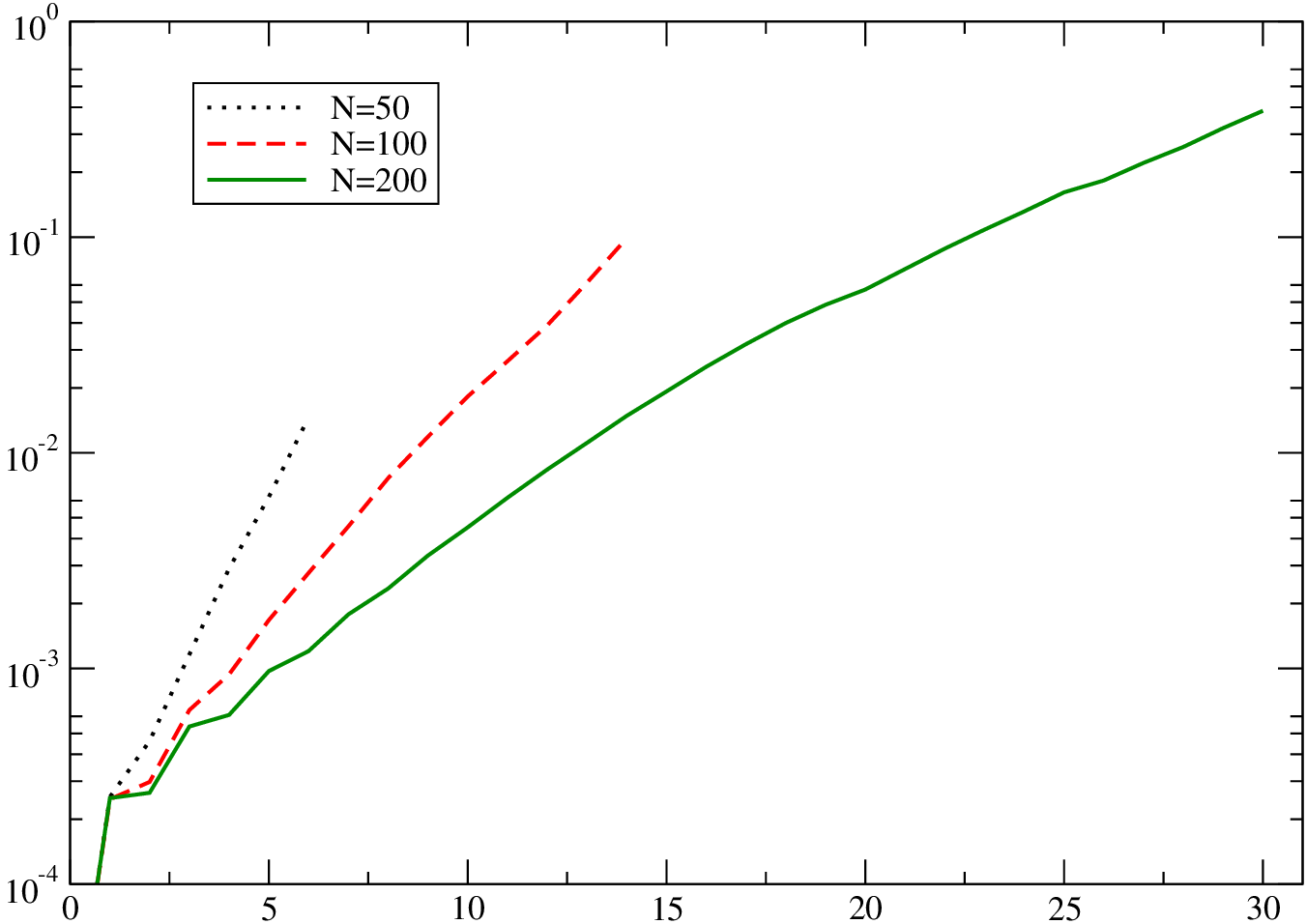}
  \caption{Log plot of the rescaled error ${\cal E}_\rho(t)$ in $g_{xx}$
  for the 1D gauge wave with shift as in Fig.~\ref{fig:GW1DSH10Dconv}
  but with 10 Neumann boundary conditions. Again the runs converge at early
  times but now the unstable mode is evident at less than 30 crossing times.}
  \label{fig:GW1DSH10Nconv}
\end{figure}

\begin{figure}[hbtp]
  \centering
  \includegraphics*[width=7cm]{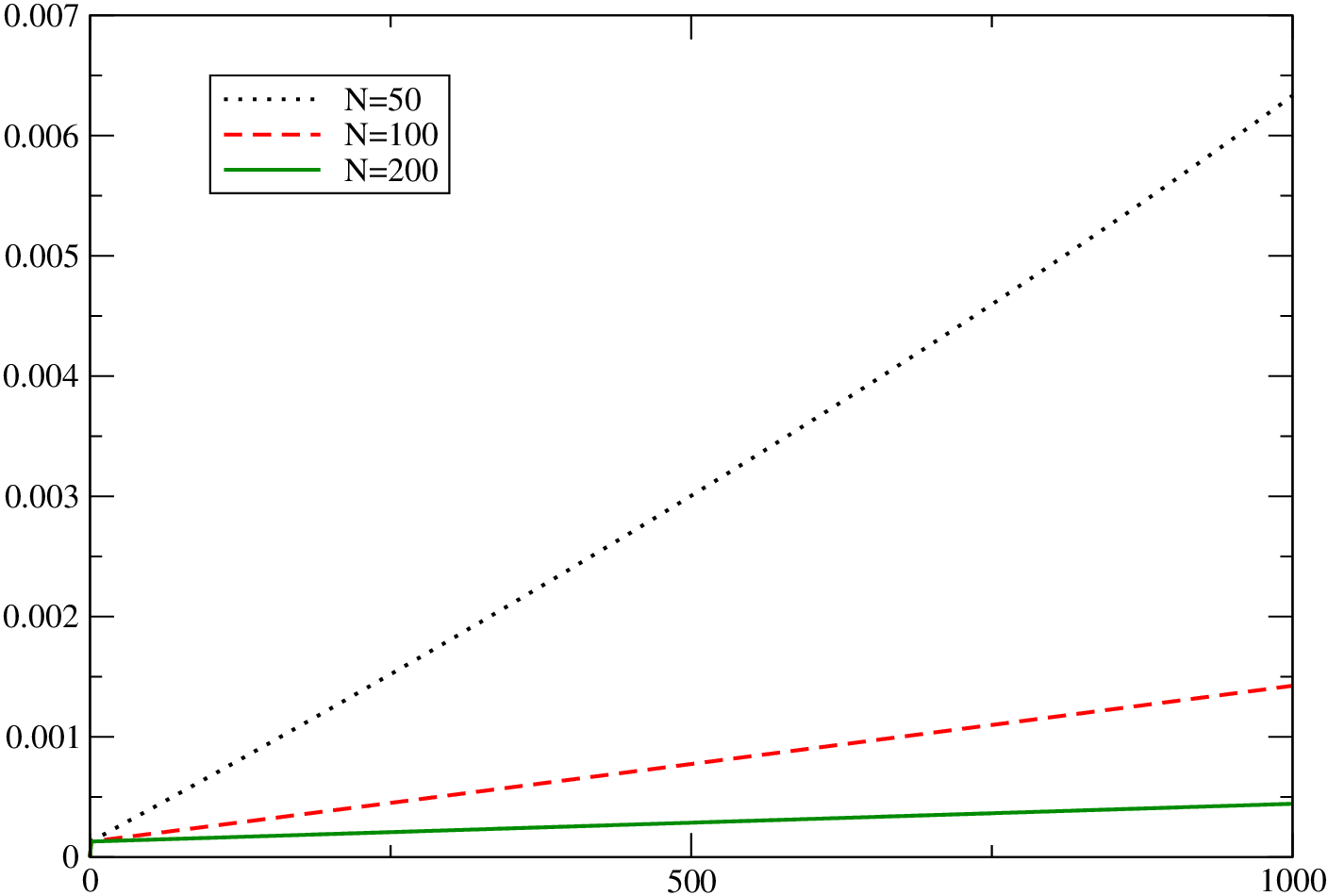}
  \caption{Plot of the rescaled error ${\cal E}_\rho(t)$ in $g_{xx}$
  for the 1D gauge wave with shift as in Fig.~\ref{fig:GW1DSH10Dconv}
  but with 10 Sommerfeld boundary conditions. The error has only a slow linear
  growth in time.}
  \label{fig:GW1DSH10Sconv}
\end{figure}

Figures \ref{fig:GW1DSH10Dconv}, \ref{fig:GW1DSH10Nconv} and 
\ref{fig:GW1DSH10Sconv} plot the rescaled error ${\cal E}_\rho(t)$ in $g_{xx}$
for shifted gauge wave tests with amplitude $A=.5$ using the bare ${\hat W}$
algorithm (no constraint adjustment or dissipation). The results in
Fig.~\ref{fig:GW1DSH10Dconv} for  10 Dirichlet boundary conditions are
much less accurate than test results for the bare algorithm reported
in~\cite{babev} for periodic boundary conditions. The periodic tests run about
4 times longer with the comparable error. The results for 10 Neumann boundary
conditions shown in Fig.~\ref{fig:GW1DSH10Nconv} are poorer yet by an additional
factor of 4 when compared to the periodic tests. These results can be
attributed to the periodic motion of the boundary introduced by the shift. As
explained in more detail in~\cite{bab}, the numerical noise can be blue shifted
as it is reflected at the moving boundaries, leading to rapid build up of
error. The results show second order convergence only at early times. For the
10-Dirichlet test we measured the convergence rate $r_{10D}(10) = 2.077$ at 10
crossing times.  For the 10-Neumann test, we measured the convergence rates
$r_{10N}(1) = 2.013$, $r_{10N}(5) = 2.760$ and $r_{10N}(10) = 3.978$, at 1, 5
and 10 crossing times. The high convergence rate at 10 crossing times is
misleading since instabilities have already dominated the $N=100$ run (the
coarser grid in the convergence calculation).

Figure \ref{fig:GW1DSH10Sconv} shows that 10 Sommerfeld boundary conditions
gives very good long term performance for the bare algorithm. The accuracy
after 1000 crossing times is much better than found in~\cite{babev} for the
bare algorithm with periodic boundary conditions, as expected since the
Sommerfeld conditions allow numerical error to leave the grid. We measured the
convergence rate
\begin{equation}
    r_{10S}(50) = 2.413
\end{equation}
at 50 crossing times.

\begin{figure}[hbtp]
  \centering
  \includegraphics*[width=7cm]{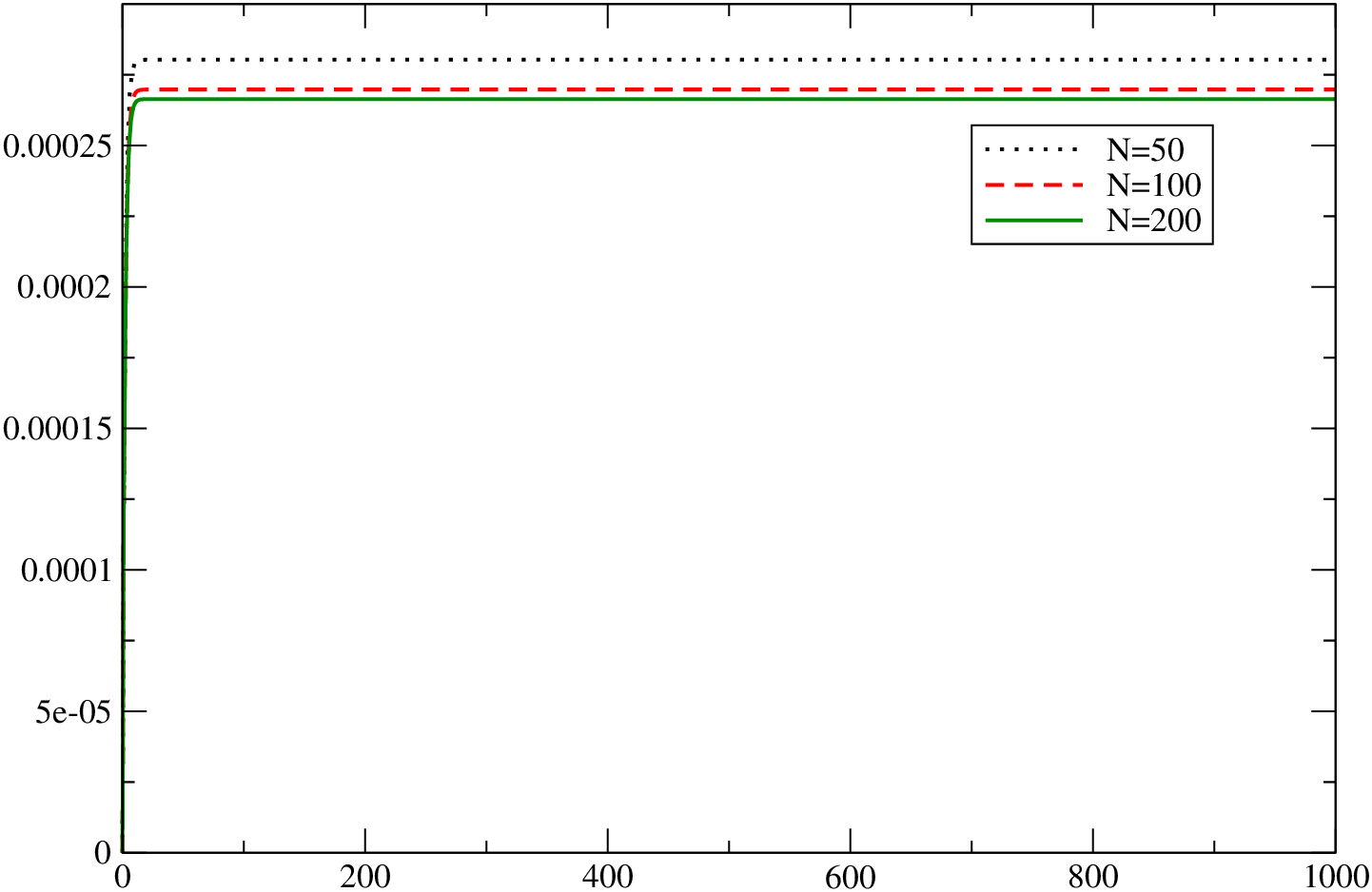}
  \caption{Plot of the rescaled error ${\cal E}_\rho(t)$ in $g_{xx}$
   for the 1D shifted gauge wave test with 10 Dirichlet boundary conditions
   as in Fig.~\ref{fig:GW1DSH10Dconv} but when the adjustment (\ref{eq:a2adj})
   is used. The dramatic improvement compared with Fig.~\ref{fig:GW1DSH10Dconv}
   is evident. There is no long term error growth during the 1000 crossing time
   run.}
  \label{fig:GW1DSH10Dconvnew}
\end{figure}

\begin{figure}[hbtp]
  \centering
  \includegraphics*[width=7cm]{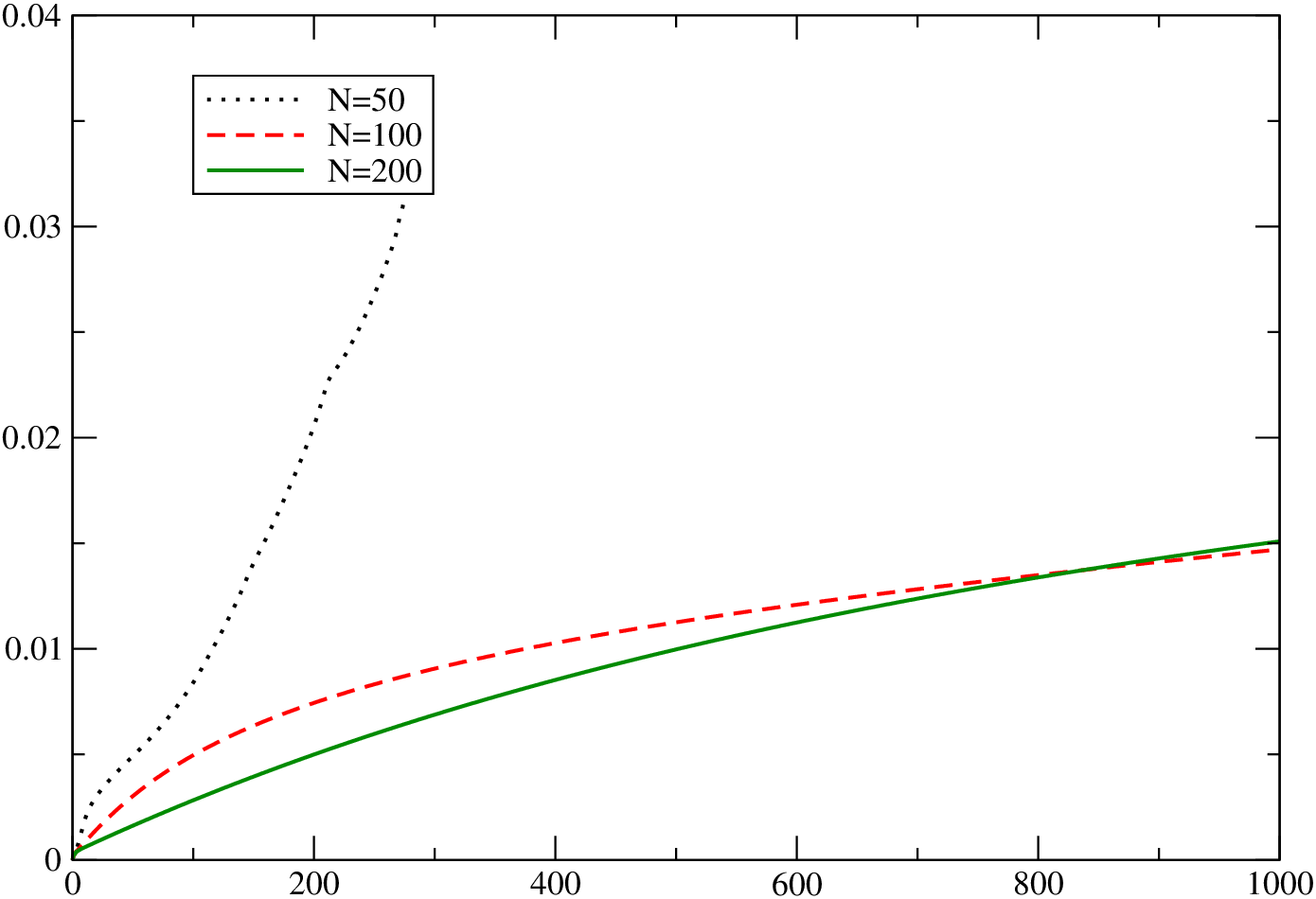}
  \caption{Plot of the rescaled error ${\cal E}_\rho(t)$ in $g_{xx}$
   for the 1D shifted gauge wave test with 10 Neumann boundary conditions
   as in Fig.~\ref{fig:GW1DSH10Nconv} but
   when the adjustment (\ref{eq:a2adj}) is used. Again, the adjustment
   leads to dramatic improvement.}
  \label{fig:GW1DSH10Nconvnew}
\end{figure}

As we have already pointed out for the case of periodic boundary conditions,
although clear second order convergence was established for short run times,
good long term performance for the shifted gauge wave tests was only possible
when constraint adjustments were used to control the unstable mode
(\ref{eq:instab}). The situation is similar in the presence of boundaries.
Figure \ref{fig:GW1DSH10Dconvnew} shows the dramatic improvement in performance
obtained by using the adjustment (\ref{eq:a2adj}), with $a_2=1$, for 1D tests
with 10 Dirichlet boundary conditions, with no dissipation added. The figure
shows good second order convergence, with only a small, non-growing error at
1000 crossing times. The convergence rate $r_{10D}(50) = 2.018$ was measured at
50 crossing times. For the same adjustment, with no dissipation, Figure
\ref{fig:GW1DSH10Nconvnew} shows the test results for 10 Neumann boundary
conditions. Again there is very good accuracy for 1000 crossing times, although
the error is larger than the Dirichlet case because the boundary conditions now
allow the instability (\ref{eq:instab}) to be excited. The convergence rate
$r_{10N}(50) = 2.877$ was measured at 50 crossing times. 

\begin{figure}[hbtp]
  \centering
  \includegraphics*[width=7cm]{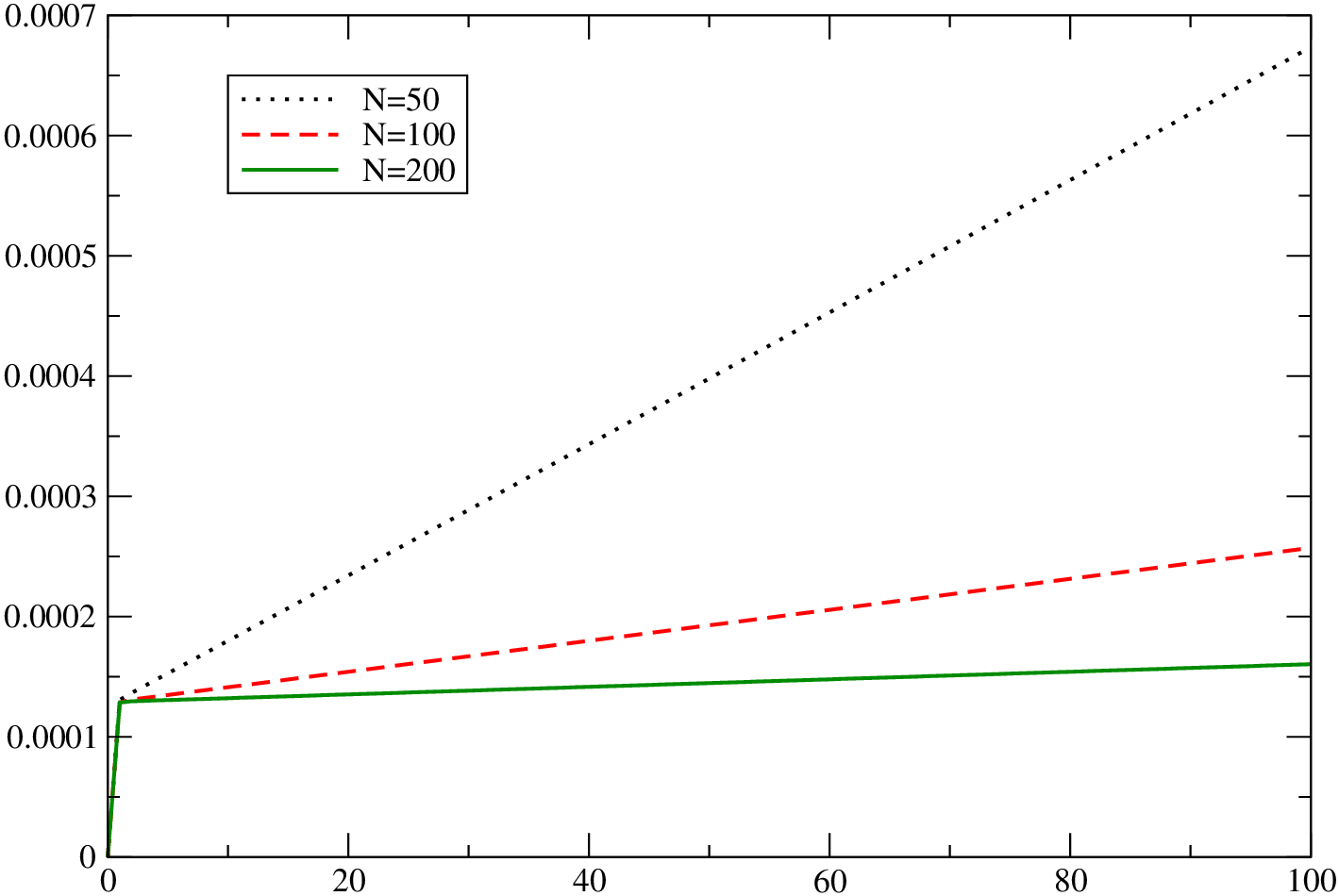}
  \caption{Plot of the rescaled error ${\cal E}_\rho(t)$ in $g_{xx}$
   with 10 Sommerfeld boundary conditions as in Fig.~\ref{fig:GW1DSH10Sconv}
   but now for the 2D shifted gauge wave test. As in the 1D test, the error has
   only a slow linear growth in time.}
  \label{fig:GW2DSH10Sconv}
\end{figure}

The 1D tests with 10 Sommerfeld boundary conditions already gave good results
for the bare algorithm which constraint adjustment or dissipation do not
substantially improve. Figure \ref{fig:GW2DSH10Sconv} shows the rescaled error
for 2D shifted gauge wave tests with the 10-Sommerfeld boundary algorithm. The
convergence rates
\begin{equation}
       r_{10S,2D}(10) = 2.096\, , \quad r_{10S,2D}(100) = 2.680\
\end{equation}
were measured at 10 and 100 crossing times.

\begin{figure}[hbtp]
  \centering
  \includegraphics*[width=7cm]{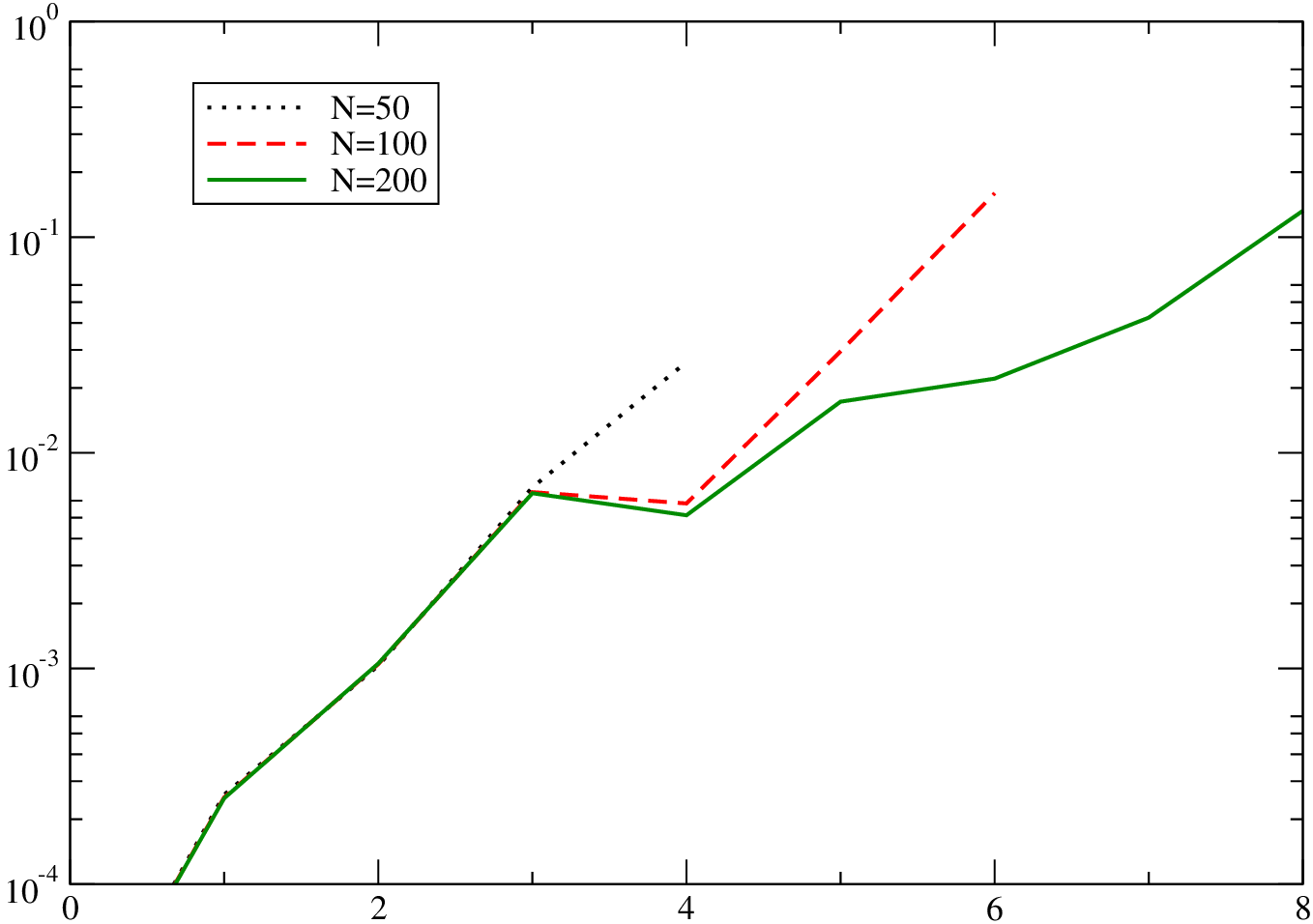}
  \caption{Log plot of the rescaled error ${\cal E}_\rho(t)$ in $g_{xx}$
   for the 1D shifted gauge wave test with 7 Neumann and 3 Dirichlet boundary 
   conditions and amplitude $A=0.5$. The instabilities which appear in less than
   8 crossing times cannot be controlled by dissipation or by the 
   constraint adjustments considered here.}
  \label{fig:AGW1DSH7N3Dconv}
\end{figure}

\begin{figure}[hbtp]
  \centering
  \includegraphics*[width=7cm]{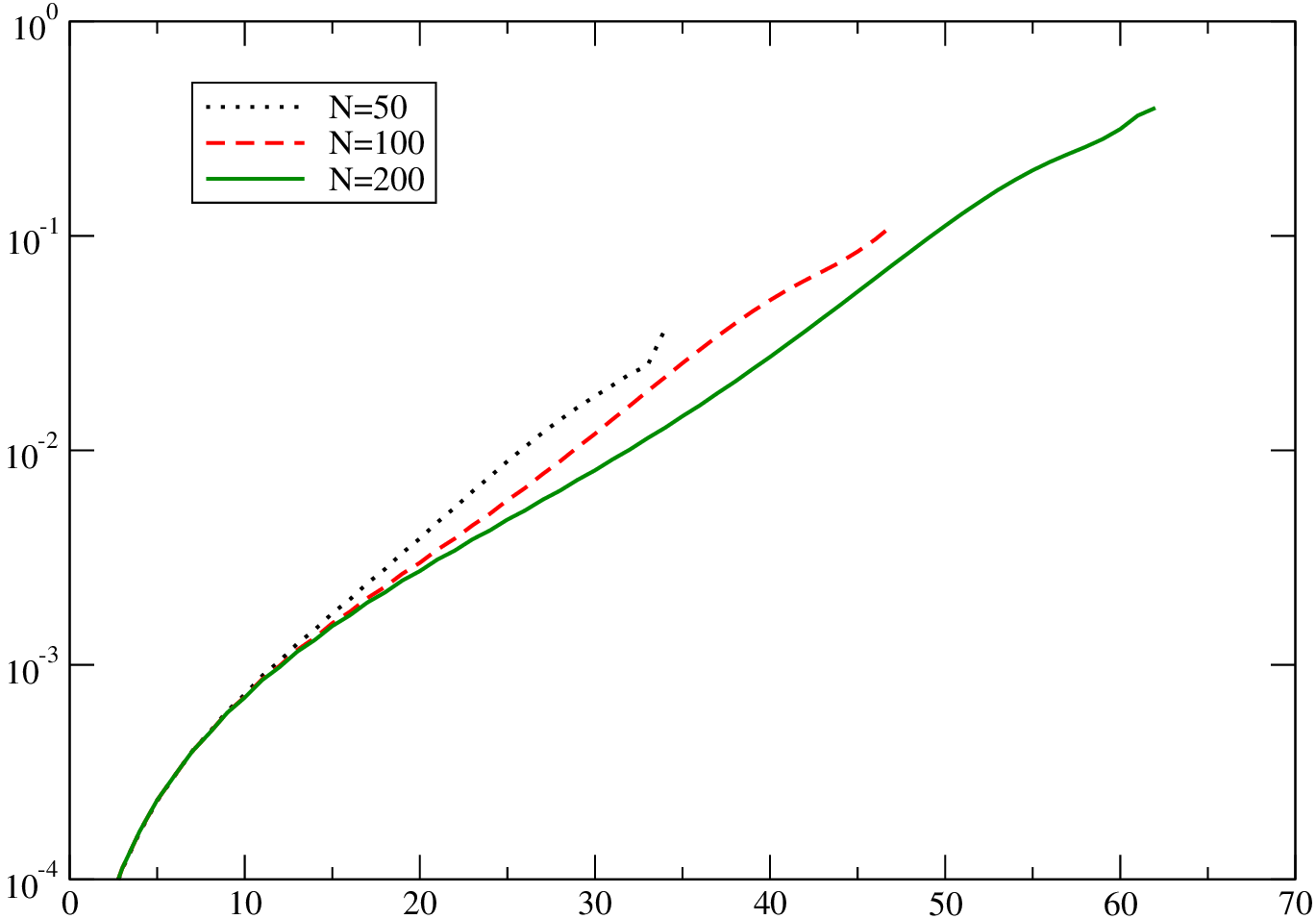}
  \caption{Log plot of the rescaled error ${\cal E}_\rho(t)$ in $g_{xx}$
   for the 1D shifted gauge wave test with 7 Neumann and 3 Dirichlet boundary 
   conditions as in Fig.~\ref{fig:AGW1DSH7N3Dconv} but with
   amplitude $A=.1$.}
  \label{fig:GW1DSH7N3DconvAm1}
\end{figure}

\begin{figure}[hbtp]
  \centering
  \includegraphics*[width=7cm]{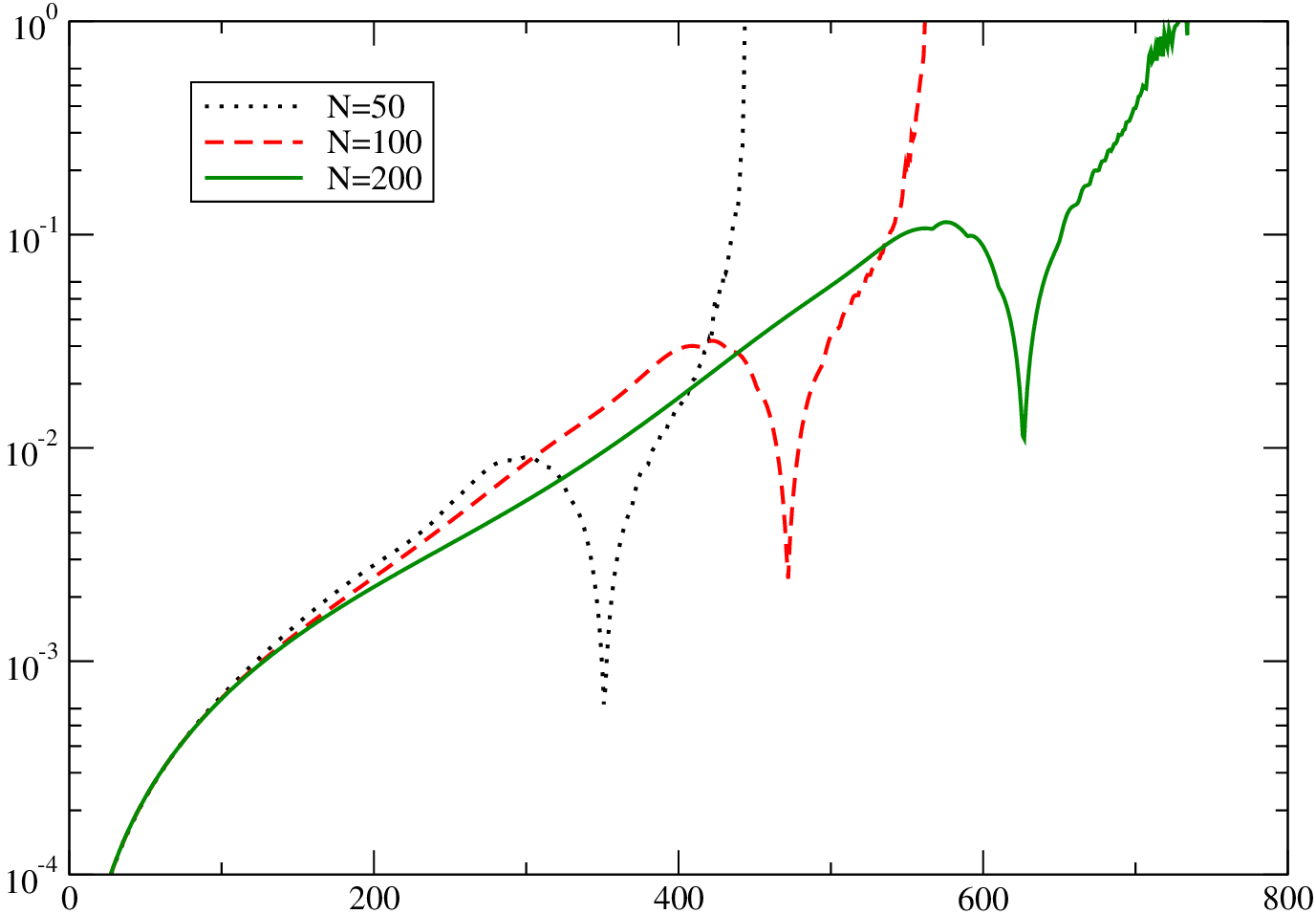}
  \caption{Log plot of the rescaled error ${\cal E}_\rho(t)$ in $g_{xx}$
   for the 1D shifted gauge wave test with 7 Neumann and 3 Dirichlet boundary 
   conditions as in Fig.~\ref{fig:AGW1DSH7N3Dconv} but with
   amplitude $A=.01$. The rapid attenuation
   of the growth rate of the instability with amplitude is evident from
   comparison with Fig's~\ref{fig:AGW1DSH7N3Dconv} and 
   \ref{fig:GW1DSH7N3DconvAm1}. }
  \label{fig:GW1DSH7N3DconvAm2}
\end{figure}

The constraint preserving boundary algorithm described in Sec.~\ref{sec:inhom}
requires 3 Dirichlet  boundary conditions (for $\gamma^{xa}$) and 7 Neumann 
boundary conditions (for $\gamma^{xx}$ and $\gamma^{ab}$). As shown in
Fig.\ref{fig:AGW1DSH7N3Dconv}, when the 3 Dirichlet and 7 Neumann pieces of
boundary data are supplied by the analytic solution the results for the shifted
gauge wave test are very poor due to the early excitation of an unstable error
mode. The rescaled error plotted in Fig.~\ref{fig:AGW1DSH7N3Dconv} shows that
good performance is maintained for less than 8 crossing times. The constraint
adjustments (\ref{eq:a1adj}) - (\ref{eq:a3adj}), as well as other adjustments
considered in~\cite{babev}, do not lead to improvement. In particular, the
adjustment (\ref{eq:a2adj}), which works remarkably well in suppressing
instabilities in the 10-Dirichlet and 10-Neumann algorithms, fails to be
effective when these algorithms are mixed. This seems related to the fact that
a Kerr-Schild pulse reflected by 10-Dirichlet or 10-Neumann conditions results
in a Kerr-Schild pulse traveling in the opposite direction. When the Dirichlet
and Neumann conditions are mixed, the reflected pulse has opposite signs for
the Neumann and Dirichlet components. As a result, the reflected pulse no
longer has the necessary Kerr-Schild properties for which the adjustment
(\ref{eq:a2adj}) was designed and a new unstable mode is excited. The effect
arises from the nonlinear coupling between the Dirichlet and Neumann components
and does not arise in the AwA gauge wave test where the Dirichlet components
vanish. The new unstable mode has long wavelength so it cannot be controlled by
dissipation and it satisfies $Q\approx  0$ (see (\ref{eq:contQ})) so that it
cannot be controlled by the semi-discrete conservation law $Q_{,t}=0$.

The effect of the constraint-preserving boundary system on the shifted gauge
wave test depends upon the performance of the underlying 3-Dirichlet, 7-Neumann
boundary algorithm so it also does not lead to good performance for amplitude
$A=.5$. For smaller amplitudes, the instability from the Dirichlet-Neumann
mixing is weaker. Figures \ref{fig:GW1DSH7N3DconvAm1} and
\ref{fig:GW1DSH7N3DconvAm2} show how the rescaled error behaves as the
amplitude is lowered to $A=.1$ and $A=.01$, when the  Dirichlet-Neumann data is
supplied analytically. For the $N=200$ grid and amplitude $A=.01$, reasonable
performance is maintained for 500 crossing times. 

\begin{figure}[hbtp]
  \centering
  \includegraphics*[width=7cm]{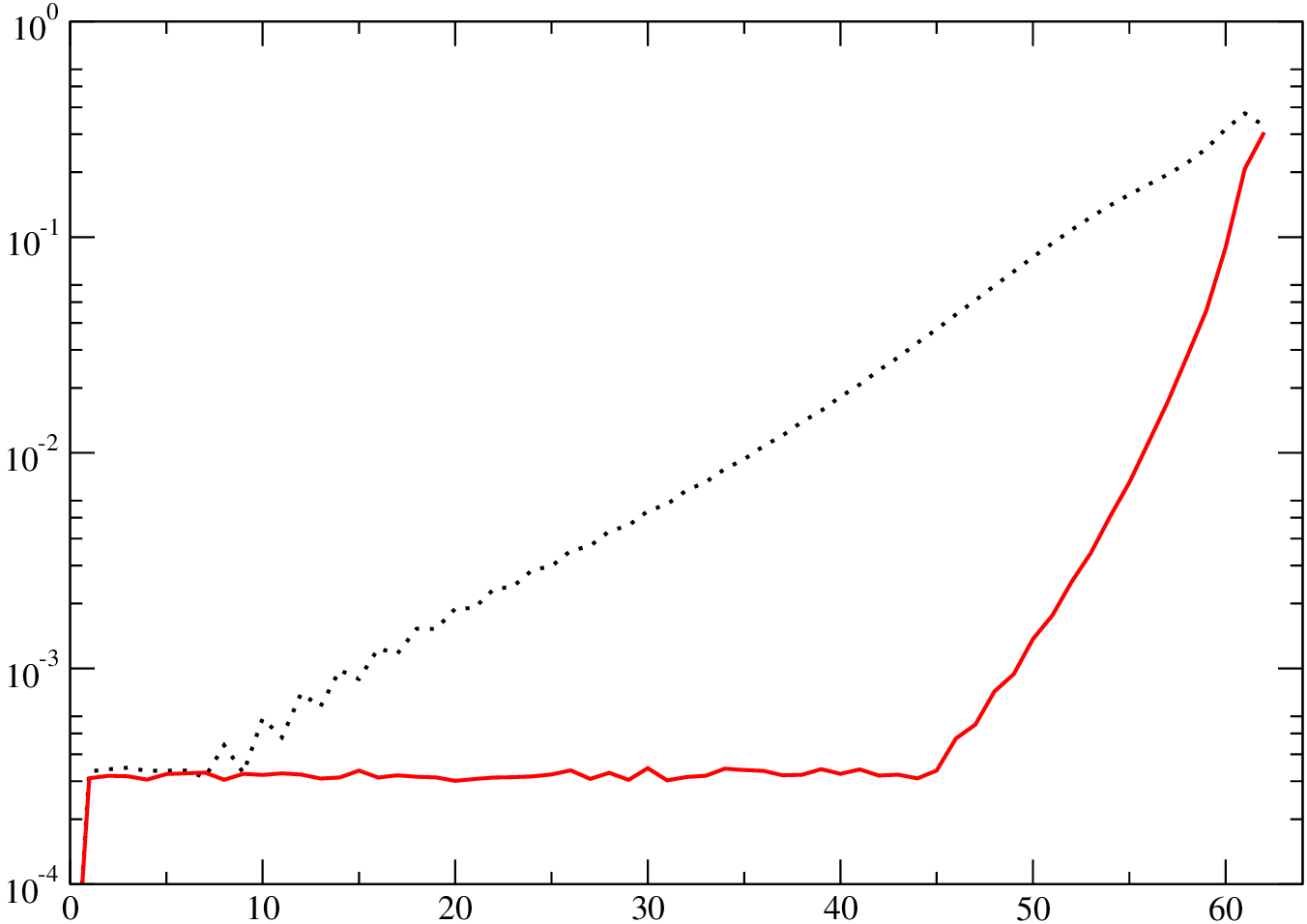}
  \caption{Log plots of the error ${\cal E}_\rho(t)$ in the constraint $C^{x}$
   for the 1D shifted gauge wave test with amplitude $A=.1$
   with 7 Neumann and 3 Dirichlet boundary conditions on an $N=200$ grid.
   The curve ($\cdot~\cdot~\cdot$) is the error when the boundary data
   is supplied analytically and the curve
   ($---$) is the error when the constraint preserving boundary system is
   applied.} 
  \label{fig:GW1DSHCP7N3DCPxAm1}
\end{figure}

\begin{figure}[hbtp]
  \centering
  \includegraphics*[width=7cm]{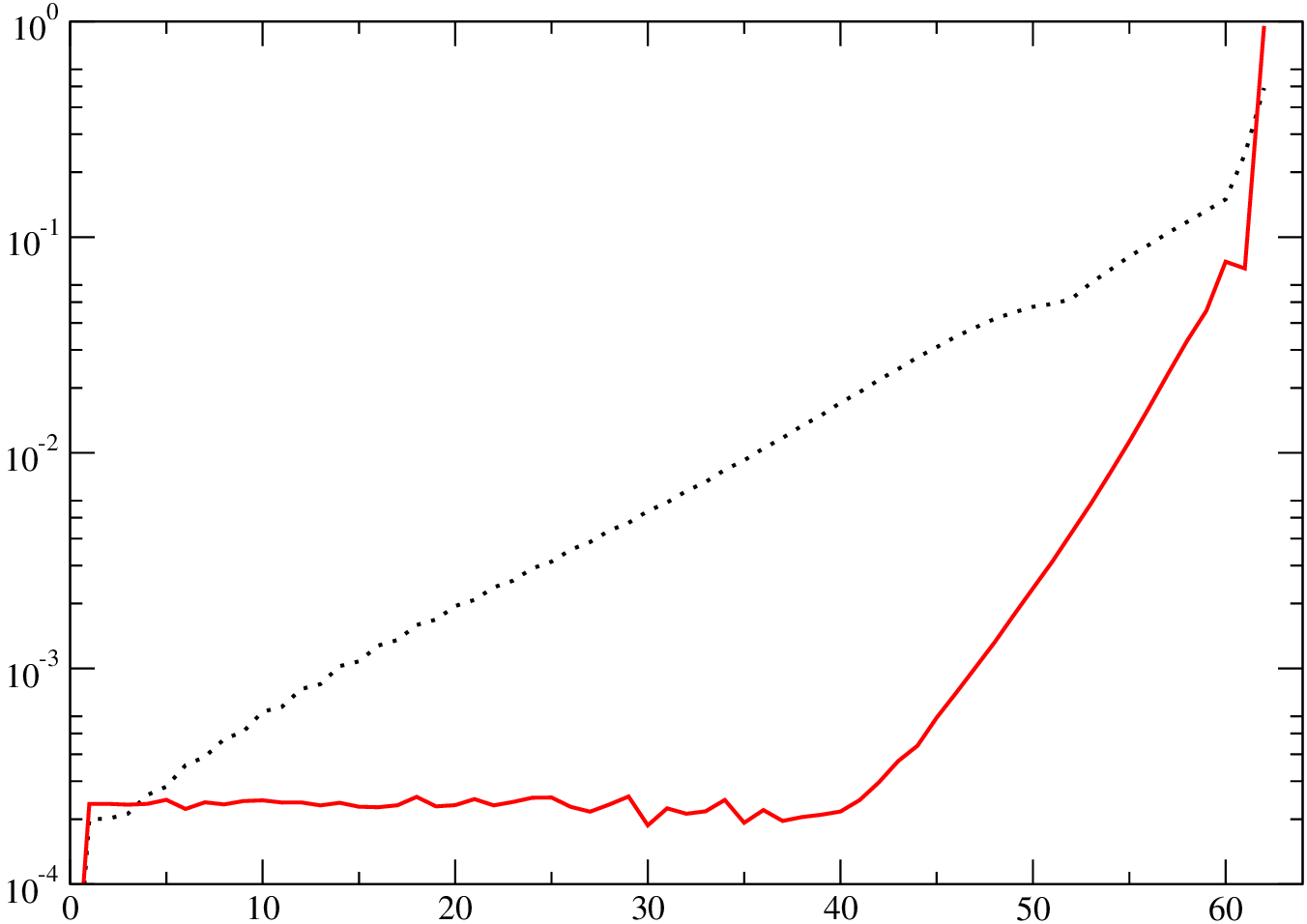}
  \caption{Log plots of the constraint error ${\cal E}_\rho(t)$ 
  as in Fig.~\ref{fig:GW1DSHCP7N3DCPxAm1} but for the component
   $C^{t}$. Although the constraint preserving boundary system ($---$)
   leads to much less constraint violation, the time scale of the unstable
   mode is unaffected.} 
  \label{fig:GW1DSHCP7N3DCPtAm1}
\end{figure}

Only slight improvement is obtained when the constraint preserving boundary
system is applied to the 3-Dirichlet, 7-Neumann algorithm. This is apparently
because the troublesome instability is a constraint preserving mode. Figures
\ref{fig:GW1DSHCP7N3DCPxAm1} and \ref{fig:GW1DSHCP7N3DCPtAm1} compare the
$\ell_\infty$ error norms for constraint violation with and without the
application the constraint preserving boundary system for the shifted gauge
wave with amplitude $A=.1$. The figures show that both the ${\cal C}^x$ and
${\cal C}^t$ components of the harmonic constraints are satisfied to much
higher accuracy when the constraint preserving boundary system is used.
However, the unstable mode still grows on the same time scale and eventually
dominates the simulation.  

\section{Summary and Implications for Cauchy-characteristic matching}

We have formulated several maximally dissipative boundary algorithms for
well-posed and constraint preserving versions of the general harmonic IBVP. The
algorithms incorporate SBP and other semi-discrete conservation laws for the
principle part of the system. In  boundary tests based upon the AwA gauge
wave,  we have demonstrated that these techniques give rise to excellent long
term (1000 crossing times), nonlinear (amplitude $A=.5$) performance for all
boundary algorithms considered. This is an especially challenging test.  A high
amplitude periodic wave enters one boundary, crosses the computational domain
and exits the other boundary. The numerical error modes, which are continously
being excited, are trapped between the boundaries by the reflecting Neumann or
Dirichlet conditions. However, this  causes no serious problem because the
chief unstable mode is suppressed by the conservation laws and the growth
of error is limited by the dissipation inherent in the system.

The shifted gauge wave boundary test proved even more challenging, as expected
from the blue shifting of the error resulting from reflection off the
oscillating boundaries. However, 10 Sommerfeld boundary conditions (on the
components of the metric) continued to give excellent long term, nonlinear
performance. Furthermore, with the use of the constraint adjustment
(\ref{eq:a2adj}), 10 Dirichlet or 10 Neumann boundary conditions also continued
to give good long term, nonlinear performance. However, nonlinear
instabilities, which did not respond to constraint adjustment, were excited by
the mixture of 3 Dirichlet and 7 Neumann boundary conditions. This led to much
poorer results and reasonable long term performance required amplitudes $A\le
.01$. When the 3-Dirichlet, 7-Neumann boundary algorithm was amended to include
the constraint preserving boundary system, the performance was only slightly
improved. The results for this test problem indicate that the mixing of
Dirichlet and Neumann boundary conditions excites a nonlinear instability of
the analytic problem which cannot be controlled by the numerical techniques
considered here.

The above results focused on the {\it mathematical} and {\it computational}
issues underlying accurate long term simulation of the IBVP but they also have
bearing on the important {\it physical} issue of providing proper boundary data
for waveform extraction by means of CCM. In the absence of analytic boundary
data, as supplied in the gauge wave testbeds, constraint preserving boundary
data must be obtained either by using a constraint preserving boundary system
or by matching to an exterior solution which accurately satisfies the
constraints. The inaccuracies which we have found in applying the constraint
preserving boundary system would only be tolerable in matching with a Cauchy
boundary in the weak field regime, i.e. far from the sources. On the other
hand, the use of 10 Sommerfeld boundary conditions shows robust long term
accuracy even in the strong field regime and would allow the economy of a
boundary close to the sources. This is a strong recommendation for a matching
algorithm based upon Sommerfeld boundary conditions, with constraint preserving
Sommerfeld data supplied by matching to an exterior characteristic solution.

Constraint preservation with a Sommerfeld boundary algorithm in CCM  is greatly
facilitated by using harmonic evolution. In CCM, there is a Cauchy evolution
region with outer boundary ${\cal B}_O$ and a characteristic evolution region,
extending to  ${\cal I}^+$, with inner boundary ${\cal B}_I$ inside the
Cauchy boundary ${\cal B}_O$. A global solution is obtained by injecting the
necessary Cauchy boundary data on ${\cal B}_O$ from the characteristic
evolution. In doing so there are two concerns: (i) injection of the Cauchy data
in the correct gauge and (ii) injection of constraint-preserving Cauchy data.
Harmonic evolution simplifies dealing with both of these concerns because the
gauge conditions and constraints reduce to wave equations for the harmonic
coordinates $x^\alpha$. By extracting data for $x^\alpha$ on the inner
characteristic boundary ${\cal B}_I$, the harmonic coordinates can be
accurately propagated  by characteristic evolution to the injection worldtube
${\cal B}_O$ as solutions of the wave equation. This supplies the necessary
Jacobian between the characteristic and Cauchy coordinates to ensure that the
above two concerns are properly handled. Furthermore, since such data is
constraint-preserving (up to numerical error), it is possible to inject the
data in (inhomogeneous) Sommerfeld form. As evidenced by the test results in
Sec.~\ref{sec:tests}, this offers a very robust way of injecting boundary data
which obeys all the constraints of the harmonic system, while avoiding the
computational error introduced by solving the boundary constraint system. This
was part of the strategy employed in~\cite{harl} in successfully implementing
long term stable CCM in linearized gravitational theory. The results of this 
paper suggest that this is also a promising approach for carrying out CCM
in the nonlinear case.

\appendix

\section{IBVP for a scalar field in a curved background spacetime}
\label{app:scalar}

The well-posedness of the IBVP for the scalar wave equation with shift can be
established by standard techniques by reducing the evolution system to
symmetric hyperbolic form. We
consider the principle part  of the scalar wave equation written in the form
\begin{equation}
         g^{\mu\nu}\partial_\mu \partial_\nu \Phi = 0.
\end{equation}
We reduce this to the first order symmetric hyperbolic system
$A^t \partial_t  u + A^i\partial_i  u = S$,
following a treatment in~\cite{cour},
by introducing the auxiliary variables    
${\cal T}=\partial_t \Phi$,  
${\cal X}=\partial_x \Phi$,  
${\cal Y}=\partial_y \Phi$,  and 
${\cal Z}=\partial_z \Phi$.
Then in terms of the 5 dimensional column vector $u={}^T(\Phi,{\cal T}, {\cal
X},{\cal Y}, {\cal Z})$, the matrices $A^\alpha$ are given by
\begin{equation}
A^t = \left( 
\begin{array}{ccc}
1 & 0 & 0 \\
0 & -g^{tt}  & 0  \\          
0 & 0 &  g^{jk}  
\end{array}
\right) 
\end{equation}
and
\begin{equation}
A^i = \left( 
\begin{array}{ccc}
0 & 0 & 0 \\
0 & -2g^{ti}  & -g^{ji}  \\          
0 & -g^{ij} &  0   
\end{array}
\right) 
\end{equation}
and $S={}^T({\cal T},0,0,0,0)$.
In this first order form, the Cauchy data consist of
$u_0=u|_{t=0}$ subject to the constraints
\begin{equation}
  ({\cal C}_x,{\cal C}_y, {\cal C}_z):=  
     ( {\cal X},{\cal Y}, {\cal Z}) -
    (\partial_x \Phi,\partial_y \Phi ,\partial_z \Phi) =0.
\label{eq:sconstr}
\end{equation}
The evolution system implies that the constraints propagate
according to
\begin{equation}
  \partial_t({\cal C}_x,{\cal C}_y, {\cal C}_z) =0. 
\end{equation}
The well-posedness of the Cauchy problem follows from the established
properties of symmetric hyperbolic systems.

We now consider the IBVP in the domain $t\ge 0$ with $x\le 0$.
The constraints propagate up the timelike boundary at $x=0$ and present
no complication.  The ``boundary matrix'' 
\begin{equation}
A^x = \left( 
\begin{array}{ccc}
0 & 0 & 0 \\
0 & -2g^{tx} & -g^{jx} \\            
0 & -g^{xj}  &  0  
\end{array}
\right) 
\end{equation}
has a 3-dimensional kernel, with a basis corresponding to the vectors
\begin{equation}
 {}^T(1,0,0,0,0), \quad
 {}^T(0,0,-g^{xy},g^{xx}, 0), \quad \mbox{and} \quad
 {}^T(0,0,-g^{xz},0 g^{xx} ) .
\end{equation}
In addition, there is one positive eigenvalue and one negative eigenvalue
\begin{equation}
 \lambda_{\pm}= \pm \lambda -g^{xt},
\end{equation}
where 
\begin{equation}
 \lambda= \sqrt{({g^{xt}})^2+\delta_{ij}g^{xi}g^{xj}} .
\end{equation}
Thus precisely one boundary condition is required.
The corresponding normalized eigenvectors are
\begin{equation}
   u_{\pm} = \frac{1}{\sqrt{ \pm 2 \; \lambda \; \lambda_\pm }} \; {}^T(0, 
    \mp \lambda +g^{xt} , g^{xx}  , g^{xy} , g^{xz} ) .
\end{equation}

The homogeneous boundary condition $Mu=0$ may be cast in the form that $u$
lies in the linear subspace $(u_+ + H u_- + u_0)$, where $u_0$ lies in the
kernel. In this subspace, the flux
\begin{eqnarray}
  {\cal F}^x=\frac{1}{2}\, {}^T(u_+ + H u_- + u_0) A^x(u_+ + H u_- + u_0)
        = \lambda_+  +\lambda_- H^2 , 
\end{eqnarray}
satisfies the dissipative condition ${\cal F}^x\ge 0$ provided
\begin{eqnarray}
    H^2 \le - {\lambda_+} \; / \; {\lambda_-} \; .
\end{eqnarray}

The homogeneous boundary condition corresponding to a choice of $H$ is
\begin{equation}
   M u: = {}^T(Hu_+ -u_-) u= 0\; .
\label{eq:bcond1}
\end{equation}
The limiting case $H = \sqrt{ - \; {\lambda_+} \; / \; {\lambda_-}}$ 
leads to the homogeneous Dirichlet boundary condition 
\begin{equation}
       \partial_t \Phi = 0
\end{equation}
 and the limiting case $H = - \sqrt{ - \; {\lambda_+} \; / \; {\lambda_-}}$ 
leads to the homogeneous Neumann boundary condition
\begin{equation}
 g^{\alpha x}\partial_\alpha \Phi = 0 .
 \label{eq:nbc}
\end{equation}

As an alternative to this standard eigenvalue analysis of the maximally
dissipative boundary condition, the simplicity of the scalar wave case allows a
direct geometrical approach. A straightforward calculation gives
\begin{equation}
    {\cal F}^x=\frac{1}{2}\, {}^T u A^x u
         =-(\partial_t \Phi) g^{x\alpha}\partial_\alpha \Phi,
\label{eq:sflux}	    
\end{equation}
which is identical to the energy flux obtained from the stress-energy tensor of
a massless scalar field. It immediately follows that ${\cal F}^x=0$ for the
above specifications of homogeneous Dirichlet and Neumann boundary conditions.
More generally, ${\cal F}^x >0$ for boundary conditions of
the form $(g^{x\mu}\partial_\mu \Phi + P\partial_t \Phi)$, with $P>0$.
An important case is the choice
\begin{equation}
      P=\sqrt{-\frac{g^{xx}}{g_{tt}}}
\end{equation}
which leads to the homogeneous Sommerfeld-like boundary condition
\begin{equation}
            L^\mu\partial_\mu \Phi=
     g^{x\mu}\partial_\mu \Phi + \sqrt{-\frac{g^{xx}}{g_{tt}}}\partial_t \Phi=0,
\end{equation}
where $L^\mu$ lies in the outgoing null direction in the
$(t^\alpha, \nabla^\alpha x)$ plane.

The well-posedness of the IBVP for the scalar wave equation with any of the
above choices of maximally dissipative boundary condition extends, by Secchi's
theorem~\cite{secchi2}, to the quasilinear reduced Einstein equations
(\ref{eq:fc}).  The extension of the homogeneous boundary condition $Mu=0$ to
include free boundary data $q$ takes the form $M(u-q) = 0$. For example, the
Neumann boundary condition (\ref{eq:nbc}) takes the inhomogeneous form
$g^{\alpha x}\partial_\alpha \Phi = q$, where the boundary data $q$ can be
freely assigned. The well-posedness of the IBVP for the wave equation with
inhomogeneous boundary data follows from the well-posedness of the homogeneous
case by a standard argument.

\section{Extrinsic curvature}
\label{app:extrinsic}

The extrinsic curvature tensor of the boundary at $x=0$ (of the domain
$x\le 0$) is given by
\begin{equation}
     K^{\mu\nu}=h^\mu_\alpha h^\nu_\beta
           \nabla^\alpha n^\beta
\end{equation}
where
\begin{equation}
    h_{\alpha}^{\mu} =\delta_{\alpha}^{\mu}-n_\alpha n^\mu
\end{equation}
is the projection tensor into the tangent space of the boundary and
$n_\alpha=(1/\sqrt{g^{xx}}) \nabla_\alpha x$ is the unit outward normal. In the
3+1 decomposition $x^\mu=(x^a,x)$, $h_{ab}$ is the intrinsic metric of the
boundary, $h_\mu^x=0$, $h_{\mu b}=g_{\mu b}$ and
$h_{xx}=g_{xx}-(1/g^{xx}$.) Also note that $h_{\cal B}=det(h_{ab})=g^{xx}g$.

We set $ n^\mu =\sqrt{g^{xx}} q^\mu =\sqrt{g^{xx}}(q^a,1)$, where
$q^a=\gamma^{xa}/\gamma^{xx}$.  In terms of the metric,
\begin{equation}
     K^{\mu\nu}= \frac{\sqrt{g^{xx}}}{2} h^\mu_\alpha h^\nu_\beta
         (g^{\alpha\rho}\partial_\rho q^\beta
	  +g^{\beta\rho}\partial_\rho q^\alpha
          -q^\rho\partial_\rho g^{\alpha\beta}).
\end{equation}

In expressing this in terms of $\gamma^{\mu\nu}$, the
identity
\begin{equation}
         \partial_\alpha g= 
             -\sqrt{-g}g_{\mu\nu} \partial_\alpha \gamma^{\mu\nu}
\end{equation}
is useful. For  example,
\begin{equation}
         \partial_x g^{\mu\nu}=  \frac {1}{\sqrt{-g}}
       ( \partial_x \gamma^{\mu\nu}- \frac{1}{2} g^{\mu\nu}
                g_{\alpha\beta} \partial_x \gamma^{\alpha\beta} ).
\end{equation}
As a result,
\begin{equation}
     K^{\mu\nu}= \frac{\sqrt{g^{xx}}}{2}( h^{\mu\rho}h^\nu_\beta
         \partial_\rho q^\beta
	  +h^\mu_\alpha h^{\nu\rho}\partial_\rho q^\alpha)
          -\frac{g^{xx}}{2\sqrt{-h}}(h^\mu_\alpha h^\nu_\beta
	  q^\rho\partial_\rho \gamma^{\alpha\beta}
	  -\frac{1}{2}h^{\mu\nu}g_{\lambda\tau} 
	         q^\rho\partial_\rho \gamma^{\lambda\tau})
\end{equation}
and
\begin{eqnarray}
     K^{\mu\nu}-h^{\mu\nu} K&=& 
       \frac{\sqrt{g^{xx}}}{2}( h^{\mu\rho} h^\nu_\beta
         \partial_\rho q^\beta
	  +h^\mu_\alpha h^{\nu\rho}\partial_\rho q^\alpha
	  -2h^{\mu\nu}h^\rho_\beta \partial_\rho q^\beta)
         \nonumber \\
           &-&\frac{g^{xx}}{2\sqrt{-h}}(
      h^\mu_\alpha h^\nu_\beta q^\rho\partial_\rho \gamma^{\alpha\beta}	  
    + h^{\mu\nu}  n_\alpha n_\beta q^\rho \partial_\rho \gamma^{\alpha\beta} ).
\label{eq:kmt}
\end{eqnarray}

The boundary version of the usual momentum
constraint implies
\begin{equation}
     h^\mu_\sigma n_\nu G^{\sigma\nu} 
            =D_\nu(K^{\mu\nu} -h^{\mu\nu} K),
\label{eq:cbmom}
\end{equation}
where $D_a$ is the covariant derivative associated with the boundary metric
$h_{ab}$.
We express this ``boundary momentum''  in the 3-dimensional form of the
$x^a=(t,x,y)$ coordinates:
\begin{equation}
     h^a_\sigma n_\nu G^{\sigma\nu} =D_b(K^{ab} -h^{ab} K)=0.
\label{eq:bmom}
\end{equation}
The identity
\begin{equation}
     \sqrt{-h_{\cal B}} D_b S_a^b = \partial_b (\sqrt{-h_{\cal B}}
       S_a^b)
      -\frac{1}{2}\sqrt{-h_{\cal B}}S^{bc}\partial_a h_{bc},
\label{eq:scd}  
\end{equation}
valid for any symmetric tensor $S^{bc}$, is useful in calculating the covariant
derivatives entering the constraint.

\begin{acknowledgments}

We thank H-O. Kreiss for many enjoyable discussions of
this material. The work was supported by the National Science
Foundation under grant PH-0244673 to the University of Pittsburgh. We used
computer time supplied by the Pittsburgh Supercomputing Center and we have
benefited from the use of the Cactus Computational Toolkit
(http://www.cactuscode.org).

\end{acknowledgments}


\begin{thebibliography}{40}

\bibitem{penrose} R. Penrose, 
``Asymptotic properties of fields and space-times'',
{\it Phys. Rev. Lett.} {\bf 10}, 66 (1963).

\bibitem{Hhprbloid}  S. Husa, 
``Numerical relativity with the conformal field equations'',
{\it Lect. Notes Phys.}, {\bf 617}, 159 (2003).

\bibitem{joerg} J.  Frauendiener, ``Conformal infinity'',
{\it Living Rev. Relativity} {\bf 7} (2004).

\bibitem{winrev} J.  Winicour, ``Characteristic evolution and matching'',
{\it Living Rev. Relativity} {\bf 8}, 10 (2005).

\bibitem{vishu} N.~T. Bishop, R.~O. G{\'{o}}mez, R.~A. Isaacson,
L. Lehner, B. Szil{\'{a}}gyi, and J. Winicour,
``Cauchy-characteristic matching'',
in {\it Black Holes, Gravitational Radiation and the Universe},
eds. B. Bhawal, B. and B.~R. Iyer, (Kluwer,Dordrecht, 1998).

\bibitem{harl} B. Szil\'{a}gyi and J. Winicour,
``Well-posed initial-boundary evolution in general relativity'', 
{\it Phys. Rev.} {\bf D68}, 041501 (2003).

\bibitem{friedrend} H. Friedrich and A.~D. Rendall, 
``The Cauchy problem for the Einstein equations'',
in {\it Einstein's Field Equations and Their Physical Implications: Selected
Essays in Honour of 
J\"{u}rgen Ehlers}, ed. B.~G. Schmidt, Springer, Berlin (2000).

\bibitem{stewartbc} J.~M. Stewart, 
``The Cauchy problem and the initial boundary value problem in
numerical relativity'', {\it Class. Quantum Grav.} {\bf 15}, 2865 (1998).

\bibitem{Friedrich98}
H.~Friedrich and G.~Nagy, {\em Commun. Math. Phys.}, {\bf 201}, 619 (1999).

\bibitem{bishbc} B. Szil\'{a}gyi, R. Gomez, N.~ T. Bishop and J. Winicour,
``Cauchy boundaries in linearized gravitational theory'',
{\it Phys.Rev.} {\bf D62} 104006 (2000). 
 
\bibitem{szilschbc} B. Szil\'{a}gyi, B.~G. Schmidt and J. Winicour,
{\it Boundary conditions in linearized harmonic gravity},
{\it Phys. Rev.} {\bf D65}, 064015 (2002). 

\bibitem{callehtigbc} C. Calabrese, L. Lehner and M. Tiglio,
``Constraint-preserving boundary conditions in numerical relativity'',  
{\it Phys. Rev.} {\bf D65}, 104031 (2002).

\bibitem{calpulreulbc} C. Calabrese, J. Pullin, O. Reula, O. Sarbach and
M. Tiglio,
``Well posed constraint-preserving boundary conditions for the linearized
Einstein equations'',
{\it Commun. Math. Phys.} {\bf 240}, 377 (2003).

\bibitem{calsarbc} C. Calabrese and O. Sarbach,
``Detecting ill posed boundary conditions in General Relativity'',
{\it Commun. Math. Phys.} {\bf 240}, 377 (2003).
{\it J. Math. Phys.} {\bf 44}, 3888 (2003).
     
\bibitem{fritgombc} S. Frittelli and R. G{\'{o}}mez,  
``Einstein boundary conditions for the 3+1 Einstein equations'',  
{\it Phys. Rev.} {\bf D68}, 044014 (2003).

\bibitem{fritgombc3} S. Frittelli and R. G{\'{o}}mez,  
``Boundary conditions for hyperbolic
formulations of the Einstein equations'',  
{\it Class. Quant. Grav. } {\bf 20}, 2739 (2003).

\bibitem{bab} M.~C. Babiuc, B. Szil\'{a}gyi and J. Winicour,
``Some mathematical problems in numerical relativity'',
{\it Lect. Notes Phys.} (to appear) gr-qc/0404092.

\bibitem{fritgombc4} S. Frittelli and R. G{\'{o}}mez,  
`Einstein boundary conditions for the Einstein equations in the
conformal-traceless decomposition`'',  
{\it Phys. Rev.} {\bf D70},064008 (2004).

\bibitem{sartig} O. Sarbach, and M. Tiglio,``Boundary conditions for
Einstein's equations: Analytical and numerical analysis'', gr-qc/0412115.

\bibitem{gunmgarcbc} C. Gundlach and J.~M. Martin-Garcia, 
``Symmetric hyperbolicity and consistent boundary conditions for second-order
Einstein equations'',  
{\it Phys. Rev.} {\bf D70}, 044032 (2004).

\bibitem{caltechbc} L.~E. Kidder, L. Lindblom, M.~A. Scheel, L.~T. Buchman
and H.~P. Pfeiffer, ``Boundary Conditions for the Einstein Evolution
System'', {\it Phys.Rev.} {\bf D71}, 064020 (2005).

\bibitem{bonabc} C. Bona, T. Ledvinka, C. Palenzuela-Luque and M. Zacek,
``Constraint-preserving boundary conditions in the Z4 Numerical Relativity
formalism'', {\it Class. Quant. Grav.} {\bf 22}, 2615 (2005).

\bibitem{multiblock} L. Lehner, O. Reula, and M. Tiglio,
``Multi-block simulations in general relativity: high order
discretizations, numerical stability, and applications'',
gr-qc/0507004.

\bibitem{calabgund} G. Calabrese and C. Gundlach, 
``Discrete boundary treatment for the shifted wave equation'',
gr-qc/0509119.

\bibitem{linsch} L. Lindblom, M.~A. Scheel, L.~E. Kidder, R. Owen
and and O. Rinne,
``A New Generalized Harmonic Evolution System'', gr-qc/0512093.

\bibitem{Choquet} Y. Foures-Bruhat,
``Theoreme d'existence pour certain systemes d'equations aux
derive\'{e}s partielles nonlinaires'',
{\it Acta Math} {\bf 88}, 141 (1952).

\bibitem{pret1} F. Pretorius,
``Numerical Relativity Using a Generalized Harmonic Decomposition'',
{\it Class. Quant. Grav.} {\bf 22}, 425 (2005).

\bibitem{pret2} F. Pretorius,
``Evolution of Binary Black Hole Spacetimes'', {\em Phys.Rev.Lett.},
{\bf 95}, 121101 (2005).

\bibitem{Fock}
V. Fock,{\it The Theory of Space, Time and Gravitation},
(MacMillan, New York, 1964).

\bibitem{wald} R.~M. Wald, ``General Relativity'',
University of Chicago Press (1984). 

\bibitem{babev} M.~C. Babiuc, B. Szil\'{a}gyi and J. Winicour,
``Testing numerical evolution with the shifted gauge wave'',
 gr-qc/0511154.

\bibitem{Friedrich} H. Friedrich, ``Hyperbolic reductions for Einstein's
equations'', {\it Class. Quant. Grav.}, {\bf 13}, 1451 (1996).

\bibitem{mex1} The AppleswithApples Alliance, M. Alcubierre et al,
``Towards standard testbeds for numerical relativity'', 
{\it Class. Quantum Grav.}, {\bf 21}, 589 (2004).

\bibitem{excis} B. Szil\'{a}gyi, H-O. Kreiss and J. Winicour, 
``Modeling the black hole excision problem'',
{\it Phys. Rev.} {\bf D71}, 104035 (2005).

\bibitem{constrdamp} C. Gundlach, J.~M. Martin-Garcia, G. Calabrese and I.
Hinder, ``Constraint damping in the Z4 formulation and harmonic gauge'', {\it
Class. Quant. Grav.} {\bf 22}, 3767 (2005).

\bibitem{alcsch} M. Alcubierre and B. Schutz, {\em J. Comput. Phys.}, {\bf
112}, 44 (1994).

\bibitem{calab} G. Calabrese, 
``Finite differencing second order systems describing black hole
spacetimes'', {\em Phys.Rev. D} {\bf 71}, 027501 (2005). 

\bibitem{live} M. Tiglio, L. Lehner and D. Neilsen,
``3D simulations of Einstein's equations:
symmetric hyperbolicity, live gauges and dynamic control of the constraints''
{\it Phys. Rev} {\bf D70}, 104018 (2004). 

\bibitem{discrenerg} L. Lehner, D. Neilsen, O. Reula and M. Tiglio,
``The discrete energy method in numerical relativity: Towards long-term
stability''
{\it Class. Quant. Grav.} {\bf 21} 5819 (2004).

\bibitem{sbp} G. Calabrese, L. Lehner, O. Reula, O. Sarbach, M. Tiglio,
``Summation by parts and dissipation for domains with excised regions'',
{\it Class. Quant. Grav.} {\bf 21}, 5735 (2004).

\bibitem{secchi2}
P. Secchi, {\em Arch. Rational Mech. Anal.}, {\bf 134}, 595 (1996).

\bibitem{cour}
R. Courant and D. Hilbert, {\em Methods of Mathematical Physics, Vol. II},
p. 594 (Interscience, NY, 1962).





\end{thebibliography}
\end{document}